\documentclass[letterpaper,twocolumn,10pt]{article}
\usepackage{usenix-2020-09}

\usepackage{shortcuts}

\begin{document}

\date{}

\title{\codename: Privilege Separating Security Monitor on RISC-V TEEs}

\author{
{\rm Mark Kuhne}\\
ETH Zurich
\and
{\rm Stavros Volos}\\
Azure Research, Microsoft
\and
{\rm Shweta Shinde}\\
ETH Zurich
} %

\maketitle
\begin{abstract}
TEE implementations on RISC-V offer an enclave abstraction by introducing a trusted component called the security monitor (SM). 
The SM performs critical tasks such as isolating enclaves from each other as well as from the OS by using privileged ISA instructions that enforce the physical memory protection.
However, the SM executes at the highest privilege layer on the platform (machine-mode) 
along side firmware that is not only large in size but also includes third-party vendor code specific to the platform. 
In this paper, we present \codename---a privilege separation approach that isolates the SM from the firmware thus reducing the attack surface on TEEs.
\codename re-purposes existing ISA features to enforce its isolation and achieves its goals without large overheads.

\end{abstract}

\pagestyle{plain}

\section{Introduction}

RISC-V is emerging as a promising ISA for upcoming platforms, ranging from high-performance computation~\cite{xuantie}, to sensors~\cite{8791364}, to accelerators~\cite{9499874}, to root-of-trust~\cite{caliptra, opentitan}.\blfootnote{Preliminary version of our upcoming Usenix Security 2025 paper.}
Given the open nature of RISC-V development model, there are several implementations of cores that adhere to the RISC-V standard. 
A RISC-V platform comprises of SoC, peripherals, and several other board components that are customized by vendors. 
In such an ecosystem, maintaining compatibility becomes a challenge and poses a threat of fragmentation. 
To this end, RISC-V standards ratify ISA specifications that the core manufacturers can follow to ensure software and compiler compatibility. 
Similarly, at a platform level, the firmware acts as an interface between the operating system (e.g., Linux kernel) and the underlying platform. 
Specifically, RISC-V offers a programmable firmware layer called the machine-mode (\mmode) that houses low-level interfaces to the CPU as well as peripherals. 
The software that runs in the firmware is sourced from different vendors that provide platform components. 
However, this opens up several security concerns. 
Any code that executes in the \mmode has direct access to all memory, including the one belonging to the OS and applications. 
Thus any bug, intentional or accidental, in third-party modules that execute in the firmware can compromise the confidentiality and integrity. 

Trusted Execution Environments (TEEs) aim to isolate application memory from privileged code executing in the OS that is prone to bugs, with an abstraction called {\em enclaves}. 
Keystone has showcased TEEs on RISC-V platforms~\cite{keystone}.
One of the main building blocks in Keystone is to use a trusted and bug-free security monitor that executes in the \mmode and creates isolated memory regions for enclaves. 
Keystone caters to the versatility of the RISC-V eco-system by assuming standard ISA feature called physical memory protection (PMP) to achieve its goals. 
Since only the \mmode has privileges to change PMP configurations, Keystone and other RISC-V TEEs trust not only the security monitor but any other \mmode software to not maliciously corrupt or tamper with PMP settings. 
However, given the nature of RISC-V firmware, any third-party code that executes in the \mmode and is part of the firmware can circumvent TEE protection.

In this paper, we aim to address this gap in the security assumptions of RISC-V TEE platforms. 
Our goal is to enforce privilege separation between the two components that execute in the \mmode: (a) the security monitor which performs critical operations of managing PMP configurations using privileged instructions; (b) the firmware which performs platform-specific tasks that usually do not use PMP-specific instructions.
This would allow platform vendors to use third-party modules in the firmware without the need for verifying that they do not break enclave isolation. 
Similarly, TEE vendors can limit their testing and verification to the security monitor without having to reason about or check the effects of the rest of the firmware components. 
To this end, we present \codename---the first system that enables privilege separation of the security monitor on standard RISC-V platforms. 

Privilege separation in general is a conceptually simple security principle which is often challenging and imperfect to enforce in real-world systems. 
To start with, practical implementations are not modular but have several dependencies and interactions between functions that need high or low-privileges. Thus, defining a partition boundary is challenging~\cite{Privtrans}. Even after addressing the question of which functionality should reside in high and low-privilege compartments, if the code has high co-dependence, it necessitates a rich interface between compartments, leading to subtle attacks (e.g., confused deputy)~\cite{civet, Hardy1988TheCD}.
Finally, the mechanisms to enforce isolation itself can be challenging. To address this, most privilege separation approaches rely on a trusted lower-layer (e.g., kernel~\cite{wedge} or hypervisor~\cite{hypersafe, sva} or hardware changes~\cite{cure}) to enforce isolation. 

Our analysis of RISC-V firmware and security monitor shows valuable insights that are conducive to privilege separation. 
Since the development of security monitors is specifically for TEEs, the implementation is fairly modular and indeed well-separated from the firmware. This is also partly because frameworks like Keystone aim to be platform agnostic to the best of their ability. 
By virtue of this, the interface between the firmware and the security monitor is also surprisingly simple---all interactions are via small set of well-defined functions that only pass data by registers without ever needing memory de-references. 
The OS and the applications also interact with the firmware necessitating interface-hardening. 
This is a standardized interface which allows the OS to interact with any platform firmware thus preserving compatibility. 
This provides a good vantage point for privilege separation to protect the monitor from untrusted OSes. 

The main challenge in realizing \codename is the privilege separation enforcement itself. Since the \mmode is already at the highest privilege, short of changing hardware~\cite{cure}, there is no other layer that can interpose the interactions between the firmware and the security monitor. 
Worse yet, since both of them execute in the \mmode, they have the privileges to always use PMP-related instructions as well as access the OS and enclave memory.
One viable approach is to employ intra-mode isolation, such that \codename only allows the security monitor to use the PMP instructions. 
However, such in-situ separation entails several challenges stemming from the nature of the low-level code that is trying to enforce isolation, in the presence of an adversary that 
can generate code, jump in the middle of instructions, trigger asynchronous events (e.g., interrupts) that change control flow, and abuse the lack of atomicity. 
These challenges have been demonstrated in prior works that take the intra-mode isolation approach for kernels and hypervisors~\cite{skee, nested-kernel}.

Our main contribution in \codename is to address the challenges of intra-mode isolation by re-purposing an existing PMP mechanism extension called {\em enhanced PMP (ePMP)}~\cite{epmp}. 
We outline four main invariants to ensure that only the security monitor can access PMP registers. Next, we make the security monitor a gateway between the OS and the firmware, ensuring secure transitions. 
Finally, we ensure that the security monitor and the firmware do not share any state that can result in control or data flow changes (memory and trap vectors).
While seemingly simple, enforcing these invariants turns out to be challenging because the security monitor has to enable and disable 
memory isolation to different physical address ranges as it transitions between the OS and the firmware.
Our careful design of interface interposition and compartment transition using PMP has to account for subtle attacks such as ROP chains that exploit PMP instructions, trap handlers that can break out of compartments, and use of PMP instructions to avoid lock-ins. 

\paragraph{Results.}
We implement \codename using ePMP emulation on two FPGA setups (Rocket and NOEL-V cores) and QEMU. We perform end-to-end benchmarking on HiFive Unmatched board (FU740 cores) with only PMP support. 
\codename does not break assumptions of OSes (Linux kernel), applications with and without enclaves (CPU and IO, databases, webservers), or firmware (OpenSBI).

\paragraph{Contributions.}
\codename is the first work that shows privilege separation of the security monitor from the firmware on a standard RISC-V platform that supports enhanced PMP. \codename evaluation on a RISC-V board, two FPGA-based cores, and an emulator shows that it does not incur drastic software changes or performance penalties. \codename is available at \url{https://dorami-riscv.github.io/dorami}.

\section{Motivation}
\label{sec:motivation}

\begin{figure}
	\centering
	\begin{subfigure}{0.37\linewidth}
		\includegraphics[width=\linewidth]{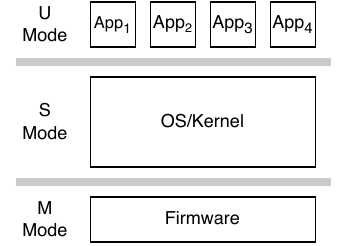}
		\caption{Standard RISC-V}
		\label{fig:setupA}
	\end{subfigure}
	\begin{subfigure}{0.29\linewidth}
		\includegraphics[width=\linewidth]{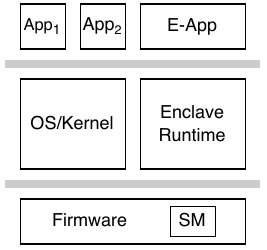}
		\caption{Keystone}
		\label{fig:setupB}
	\end{subfigure}
	\begin{subfigure}{0.29\linewidth}
	        \includegraphics[width=\linewidth]{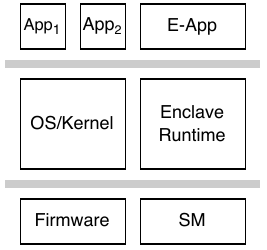}
	        \caption{\codename}
	        \label{fig:setupC}
         \end{subfigure}
	\caption{Software stack on RISC-V systems. 
 }
	\label{fig:setup}
\end{figure}

RISC-V has three execution modes as shown in Fig.~\ref{fig:setup}.
The machine-mode (\mmode) houses the firmware which runs at the highest privilege.
It has access to all memory regions and the privilege to change machine registers (CSRs equivalent to control registers on x86\_64 and special registers on Arm)~\cite{riscvSpecs, ARMregisters, intelsoftwaredev}. 
The supervisor-mode (\smode) has lower privileges than the \mmode and houses the OS.
Applications execute in user-mode (\umode) and the OS isolates them from each other using page tables. 
Since the \smode can access any \umode memory, an attacker that can execute its own applications can exploit bugs in the OS to bypass process isolation to compromise other applications (Fig.~\ref{fig:setupA}).
To reduce this attack surface, RISC-V TEEs use hardware primitives to 
create an {\em enclave} abstraction, wherein the OS executing in \smode and untrusted applications executing in \umode cannot access enclave memory~\cite{sanctum, keystone, penglai, cure, elasticlave}.
In particular, one approach is to use physical memory protection feature (PMP)---standard feature in RISC-V privilege specification supported on most 64-bit platforms~\cite{u54, u74, xuantie, ariane}---to prohibit the OS from accessing parts of physical memory. 
Fig.~\ref{fig:setupB} shows one such TEE based on Keystone. It 
introduces a trusted security monitor in \mmode who is in charge of managing PMP registers to create contiguous physical memory regions for enclaves that are inaccessible to the OS. Keystone can execute security-sensitive applications and supporting runtimes in the enclaves to achieve its security goals.

\begin{table}
\centering
\setlength{\tabcolsep}{1.3pt}
\caption{RISC-V Firmware Analysis.}
\label{tab:firmware-analysis}
\resizebox{\columnwidth}{!}{%
\begin{tabular}{lcccclr}
\toprule
\textbf{Firmware} & \textbf{Version} &\textbf{Date} & \textbf{Modules} & \textbf{Lang} & \textbf{Supported Platforms} & \textbf{LoC} \\
\midrule
OpenSBI & 1.3 & 10.2023 & 8 & C & All modern RISC-V boards  & 23,942 \\
BBL & 1.0 & 08.2019 & 0 & C & Early RISC-V boards  & 23,060 \\
RustSBI & 0.3.2 & 10.2023 & 4 & Rust & Qemu, Unmatched, Allwinner D1*, K210  & --\footnotemark \\
OreBoot & 6b9ddbe  & 10.2023 & 8 & Rust & Qemu, Unleashed, Allwinner D1*  & 23,552 \\
\bottomrule
\end{tabular}
}
\end{table}

\subsection{Firmware vs. Security Monitor}
\label{ssec:fw-sm}

On closer examination, all RISC-V TEEs rely on a security monitor---a trusted component executing in the \mmode. 
For example, the Keystone security monitor isolates enclaves (code and data), generates attestation reports, facilitates secure context switches between the OS and the enclave, and securely destroys enclaves before the OS can reclaim the physical memory. 
Other TEEs have security monitors that perform similar protections (e.g., shared memory, intra-enclave isolation, peripheral isolation).
The monitors typically have small codebases (10-15\,KLoC), can be programmed in memory-safe languages~\cite{ace-riscv}, and are subject to formal verification~\cite{Ozga_2023}.

Other than the security monitor, the \mmode also houses the firmware.
The exact functionality and codebase comprising the firmware depend on the specifics of the platform (the components on the SoC, the board, vendor-specifics, device management, interrupt controller, timer, IO, synchronization). 
To allow diversity of such platforms while providing a uniform interface to the OSes, RISC-V specifies a Supervisor Binary Interface (SBI) that the firmware implements and the S-mode assumes~\cite{sbiSpec}.
We analyzed 4 open-source firmware projects: berkeley bootloader (bbl), OpenSBI, RustSBI,\footnotetext{RustSBI LoC is not precise 
since its build process is platform-dependent.}
and oreboot~\cite{bbl, opensbi, rustsbi, oreboot}, summarized in Tab.~\ref{tab:firmware-analysis}.
We observe that the C implementations are capable of booting Linux, but are prone to memory vulnerabilities. 
While the Rust implementations aim to address this gap, they have not reached a maturity to support fully functional Linux.
More importantly, the firmware comprises not only the core-functionality (e.g., firmware versioning) but also third party code specific to the components (e.g., cache controller, timer) on the RISC-V platform. 
For example, OpenSBI has plugins from 8 vendors. 
Since it executes in \mmode with the highest privileges, if there are any bugs in the firmware (accidental or intentionally introduced by third-party drivers), the attacker can exploit them to corrupt the firmware.
For example, prior works have shown several bugs in firmware from devices such as IoT sensors, to phones, to accelerators~\cite{fie-usenix13,p2im,fuzzware,sok-tz,el3xir}.
The RISC-V firmware eco-system is not yet mature (e.g., no CVEs, security bulletins, advisories), so we cannot report the number of bugs in each of the implementations listed in Tab.~\ref{tab:firmware-analysis}.
We investigated OpenSBI bugs manually by
sampling the RISC-V mailing list.
We found several bug reports and fixes for buffer overflows~\cite{005078}, missing null termination~\cite{005170}, NULL pointer dereference~\cite{007184}, missing checks~\cite{007050}, and incorrect masks~\cite{007147}.
Thus, the security monitor in \mmode can be a target of firmware exploits to bypass enclave isolation.

\subsection{Challenges in Privilege Separation}
\label{ssec:challenges}

Ideally, by the principle of least privilege, only the security monitor should be able to change critical information pertaining to enclave isolation. 
Similar to OS designs that advocate for kernel's privilege
separation, such principled separation of the SM can significantly reduce not only the TCB but also the effective attack surface.
To draw a further analogy to OS design, the kernel must not access user memory in order to protect itself from exploits where the attacker tricks the kernel into accessing attacker-controlled user application memory. For example, on x86\_64, the kernel in ring 0 is denied from accessing ring 3 memory; instead, it has to explicitly use SMAP/SMEP~\cite{intelsoftwaredev} features to enable the access. This precaution limits the attacker's capabilities when exploiting kernel vulnerabilities.
On RISC-V, the security monitor or, for that matter, any \mmode code does not need to access any \sumode memory unless explicitly required (e.g., when the OS requests the SM to generate an attestation measurement of an enclave via an ecall). 
Thus, the SM should be disallowed from accessing \sumode memory by default.

Partitioning the security monitor and the firmware in two compartments is a worthy goal if one can achieve three requirements: the boundary of the partitioning, the interface between the partitions, and the enforcement mechanisms for the partitioning.
The first step is to decide which functionality of the \sm and the firmware should execute with the higher-privilege. From our analysis of several open-source security monitors for RISC-V TEEs~\cite{sanctum, keystone, elasticlave, penglai}, we observe that the monitor typically has a clean and clear separation---it handles enclave lifecycle and accesses certain hardware interfaces, it seldom interacts with the firmware for servicing enclaves. On the other hand, the OS invokes the firmware quite often but it almost always does so for non-enclave operations.
Due to such modular nature, deciding the partition boundary is a relatively easy task, especially when compared to similar efforts for monolithic kernels and applications~\cite{nested-kernel}. 
The second step is to ensure that the privilege separated compartments have a minimal interface, that does not require excessive data transfers, thus minimizing the attack surface further.
Our analysis shows that the \sm and the firmware do not need to pass any data beyond register values, all of which are non-pointers. 
Tab.~\ref{tab:api-analysis} shows the summary of our interface analysis for four open-source RISC-V based TEEs.

\begin{table}[]
\centering
\caption{Number of APIs from the \smode (OS) to security monitors (P), from P to firmware (F).}
\label{tab:api-analysis}
\resizebox{\columnwidth}{!}{%
\begin{tabular}{@{}llllrrr@{}}
\toprule
\textbf{Monitor} & \textbf{Release} & \textbf{Version} & \textbf{Firmware} & \textbf{P LoC} & \textbf{P} & \textbf{F} \\ \midrule
Keystone~\cite{keystone}       & Mar 2021 & {\color[HTML]{000000} 1.0} & OpenSBI     & 8,401 & 10 & 11 \\
Elasticlave~\cite{elasticlave} & Sep 2023 & {\color[HTML]{000000} 1.0} & BBL         & 8,538 & 22 & 5 \\
Penglai~\cite{penglai}         & Aug 2023 & {\color[HTML]{000000} Tvm} & BBL         & 10,232 & 19 & 6 \\
Sanctum~\cite{sanctum}         & Feb 2020 & {\color[HTML]{000000} 1.0} & independent & 5,092 & 31 & N/A \\ \bottomrule
\end{tabular}%
}
\end{table}

The final step in achieving such privilege separation is enforcing the isolation. 
For example, one approach is to introduce a lower layer that transparently isolates two components (e.g., using a hypervisor to do intra-kernel isolation~\cite{skee}).
Since \mmode is the lowest level that can execute software on RISC-V, the only option is to rely on the micro-architecture which would require, perhaps undesirable, platform changes~\cite{cure}. 
Alternatively, prior approaches achieve in-situ isolation by using hardware features~\cite{nested-kernel,skee,hilps,nexen,elisa} and instrumentation~\cite{xfi}.
Either way, the enforcement must ensure that the attacker can never misuse the isolation-specific operations. 
For example, when the kernel uses SMAP/SMEP, it has code snippets that conditionally allow it to access user memory. 
If the kernel has memory bugs, the attacker can do a ROP-chain attack to first use the SMAP/SMEP features~\cite{smepbypass}.
As another example, when the hypervisor isolates the kernel, it has to mediate all interfaces to correctly enforce the isolation. 
Finally, the design has to consider low-level operations such as interrupts, exceptions, and coarse-grained privileges.

\subsection{Problem Statement}
\label{sect:problem_statement}

\codename aims to ensure that the monitor is privileged separated from the firmware. 
It introduces the notion of a \textit{\primary (P)} and a \textit{\secondary (F)}, where the \primary is strictly more privileged than the \secondary. 
\compi houses the \sm and does security-critical operations (e.g., enclave lifecycle) that requires access to critical hardware instructions (e.g., PMP management, trap vectors).
\compii houses the remaining \mmode code, prominently the platform firmware.

The OS, applications, and enclaves can invoke services from the \mmode by explicitly invoking \textit{ecalls} to synchronously switch to \mmode. 
Additionally, if the CPU core receives a runtime exception (divide-by-zero) or interrupt (timer), the execution switches asynchronously to \mmode handlers. 
\codename has to ensure that any such interfaces from the \sumode to the \mmode are guarded. Further, \codename has to route the request to the correct compartment. For example \codename 
has to invoke the \secondary if the \sumode wants to invoke firmware functions and \primary if the OS wants to create a new enclave.

Since the \mmode is the most privileged layer on the platform, \codename has to rely on hardware for isolation. Specifically, \codename uses the RISC-V hardware's ability to perform fast physical memory isolation via PMP configurations. 
However, RISC-V allows any \mmode code to change PMP configurations. So \codename has to enforce intra-mode isolation to prohibit the \secondary from tampering the PMP configurations. 
To this end, \codename must enforce four security invariants:

\noindent 
\invepmp: {\textit{Only P is trusted to configure and change PMP regions.}}
\codename only allows P to access PMP registers,\footnote{Barring one case that we will explain in Section~\ref{sec:s2p}.} since the isolation itself is done using PMP. 

\noindent 
\invenex: {\textit{P is the only entry and exit point between \sumode and \mmode.}}
\codename introduces new transitions for existing interface between the \sumode and \mmode to ensure isolation---not only between compartments but also between the modes. This includes synchronous interfaces (ecalls) and asynchronous transitions (traps). 

\noindent 
\invintf: {\textit{F can only invoke P via a fixed interface.}}
\codename provides only one interface from P to enter F, without nesting. 
Such explicit interface allows \codename to ensure a fixed entry and exit point to and from the \secondary to enforce PMP isolation as well as control the data that is passed between the compartments.

\noindent 
\invmt: {\textit{P and F do not share any memory regions or trap vectors.}}
\codename allocates independent physical memory regions to the compartments.
It also maintains different interrupt vector tables (IVTs, referred to as trap vectors on RISC-V) per compartment. 
This way, if F installs malicious handlers that are triggered when P is executing, they are not executed. 
\codename performs a secure context save and restore across compartments and modes during transitions.

\paragraph{Threat Model \& Scope.}
The \primary is trusted by all components on the systems and is assumed to be bug-free.
The \primary does not trust the \secondary and any other code executing in \sumode (OS, user apps, enclaves), and hence avoids accessing any memory belonging to these untrusted components. 
The \secondary does not trust the \sumode, as is typical on RISC-V platforms.
The enclaves do not trust each other or the OS and the OS does not trust the enclaves, as is typical for TEEs. 
The attacker has full control of the host OS, can launch malicious enclaves, and exploit bugs in the firmware with the goals a) executing arbitrary code in firmware; b) leak code, leak/corrupt data of enclaves.
The attacker can use the \secondary to compromise the \primary who is in charge of isolating the compartments and enclaves. 

\codename does not address DoS, side-channel attacks, and hardware bugs that may break the TEE guarantees provided to enclaves.
It relies on existing mechanisms that provide protection against these attacks.
\codename does not protect against a \secondary that can launch DoS attacks against the \primary.

\subsection{Existing Approaches}
\label{sec:existing-approaches}

Tab.~\ref{tab:related-compare} shows prior works that enforce isolation on different architectures and layers.
\codename shares challenges and approaches (e.g., using call gates, binary scanning, and interrupt handler for sanitization) with them. 
In our experience, these are common for any intra-mode isolation that does not use instrumentation~\cite{skee, nexen, elisa}.
We compare \codename to two approaches in particular for achieving intra-mode isolation by either modifying the hardware or by repurposing standardized hardware features. 
Cure~\cite{cure} provides intra-mode isolation and exclusive assignment of peripheral devices to enclaves. But it modifies hardware to add CPU and bus-level checks: CPU sets enclave ID and system bus does the access control. In \codename, instead, our goal is to provide intra-mode isolation in the most privileged layer while relying on standardized hardware features (i.e., ePMP) and without any additional changes to hardware. Our work shows how to overcome challenges arising from this goal (discussed in Section~\ref{sec:insuff-pmp}).

Nested Kernel (NK)~\cite{nested-kernel} isolates memory management functions in kernel space by protecting the page table translation configuration to create two compartments: nested kernel with higher privilege to perform memory reconfigurations and outer kernel that serves the remaining kernel functions.
Comparing NK to \codename, we observe some similarities in this isolation principle, however the insights are different.
First, NK operates in kernel space (Ring 0) and uses MMU and write-protect (WP) bit. This combination ensures that the attacker does not jump to the privilege nested kernel code or data.
On the other hand, \codename operates in firmware space (\mmode) where it cannot use page-table based isolation with the MMU and there is no WP-bit; instead it has to resort to \epmp. Because of this difference, \codename has to address the challenge of an attacker jumping to code gadgets in a different way---instead of unmapping the pages, we reconfigure the PMP to make the gadgets inaccessible.
While this may seem a minor difference, \codename has to introduce a PMP reconfiguration code snippet which can then in turn be used as a gadget. 
To make this gadget useless to an attacker, \codename uses a novel PMP trick---as soon as the attacker tries to misuse the gadget we lock the attacker's memory.
Next, NK reasons about interrupts by changing all the interrupt handler code to check that the WP-bit is set. This ensures that the outer kernel cannot misuse the handlers to access the nested kernel.
In contrast, \codename allows the firmware to use its own arbitrary interrupt handlers. It simply ensures that the executing firmware can never change its own memory permissions. Specifically, \codename uses binary scanning, non-writeable code pages, and atomic compartment switching.

\begin{table}[]
\centering
\caption{Related Work Summary. Comparison of prior works based on hardware modifications, standardized hardware features, target ISA, privilege level for the trusted protector that enforces isolation, isolation abstractions used for the protectee, isolation enforcement type, and TCB size.}
\label{tab:related-compare}
\resizebox{\columnwidth}{!}{%
\begin{tabular}{@{}lclllllr@{}}
\toprule
\multicolumn{1}{c}{\textbf{Approach}} &
  \multicolumn{1}{c}{\textbf{\begin{tabular}[c]{@{}c@{}}HW\\Mod\end{tabular}}} &
  \multicolumn{1}{c}{\textbf{\begin{tabular}[c]{@{}c@{}}Std.\\Feature\end{tabular}}} &
  \textbf{ISA} &
  \multicolumn{1}{c}{\textbf{\begin{tabular}[c]{@{}c@{}}Protector\\(Trust Priv.)\end{tabular}}}  &
  \multicolumn{1}{c}{\textbf{\begin{tabular}[c]{@{}c@{}}Protectee\\(Isolation Abs.)\end{tabular}}}  &
  \multicolumn{1}{c}{\textbf{\begin{tabular}[c]{@{}c@{}}Iso.\\Type\end{tabular}}}  &
  \multicolumn{1}{c}{\textbf{\begin{tabular}[c]{@{}c@{}}TCB\\ KLoC\end{tabular}}} \\ \midrule
Cure~\cite{cure}        & \cmark & Bus Filter   & RISC-V  & FW         & FW, OS, Enclave & ex-situ & ~3.0        \\
NK~\cite{nested-kernel} & \xmark & MMU, WP      & x86     & Kernel     & Kernel            & in-situ & ~4.0        \\
SKEE~\cite{skee}        & \xmark & MMU          & ARMv7/8 & Hypervisor     & Kernel & in-situ & Hyp. \\
NeXen~\cite{nexen}      & \xmark & PT nesting   & x86     & Hypervisor & Monitor, Domains  & in-situ & N/A    \\
Elisa~\cite{elisa}      & \xmark & EPT          & x86     & Kernel/VM  & VM                & in-situ & 1.4           \\
SVA~\cite{sva}          & \xmark & LLVM         & x86     & SVM        & OS, Services & Instr.  & 5.1           \\
\textbf{\codename}      & \xmark & ePMP         & RISC-V  & Monitor    & Monitor, Firmware & in-situ & 10.6      \\ \bottomrule
\end{tabular}%
}
\end{table}

\section{Rationale for using ePMP}
\label{sec:insuff-pmp}

All prior RISC-V TEEs use a standard ISA feature called physical memory protection (PMP).
Although PMP can effectively isolate enclaves and the host OS from each other, it has several limitations that make it unsuitable for realizing compartmentalization within the M-mode.
First, we explain the details of this PMP-based isolation and outline why it is not a good fit for \codename.
Then we introduce another standardized RISC-V ISA feature called ePMP (stands for enhanced PMP) and explain how it can realise \codename.\footnote{The RISC-V specification refers to the ePMP extension as Smepmp.}

\paragraph{Background: Physical Memory Protection (PMP).} 
It is a feature for RISC-V CPUs that can be used to divide the physical memory into {\em PMP regions}, where a region is defined by a range of continuous physical addresses.
Each of these regions can then be associated with specific access permissions defined by a PMP configuration register.
RISC-V introduces two new types of registers, \verb|pmpaddr| and \verb|pmpcfg|, to enable PMP regions and region-specific configurations, respectively. 
Since these are privileged registers (CSRs), only \mmode is allowed to change them. 
Once PMP regions are set up, on each memory access originating from \sumode, 
the hardware checks if the target address is protected by a PMP region. 
If so, it further checks whether the access type (R/W/X) is permitted according to the PMP configuration.

\begin{figure}[]
    \centering
    \includegraphics[width=1\columnwidth]{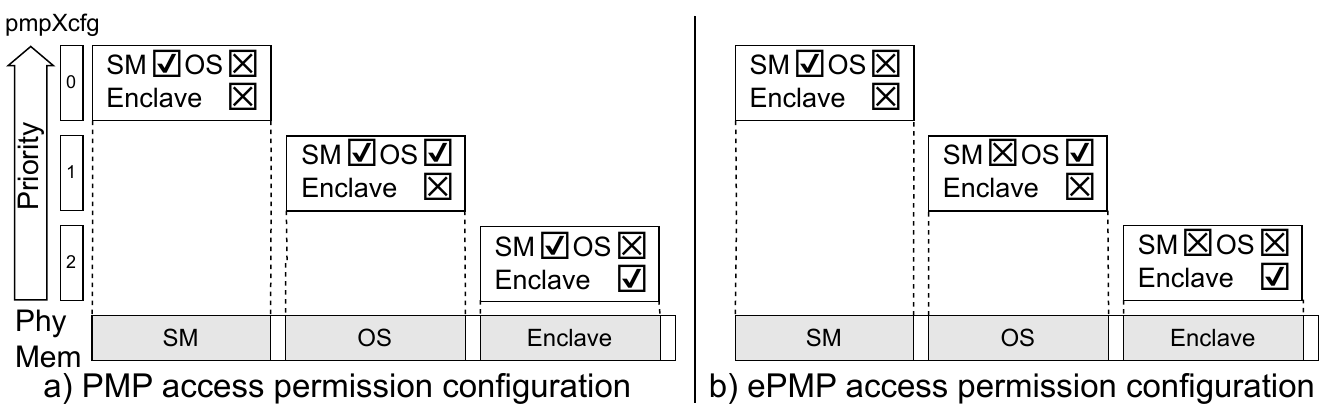} 
    \caption{The system memory is divided into 3 memory regions, one each for SM, host-OS, and enclave. With \textbf{PMP}-based isolation, host-OS and enclave cannot access each others region, but the SM always has access to both, regardless of the configuration in the \texttt{pmpXcfg} registers, where \texttt{X} is a placeholder for a specific entry. With \textbf{ePMP}-based isolation, the SM can also only access the memory in its own region.}
    \label{fig:pmp-deployment}
\end{figure}

PMP can be used to isolate \mmode from \sumode, such that the hardware denies any attempts by unprivileged software (e.g., OS) to access M-mode memory (e.g., preventing modifications of privileged system components).
Specifically, \mmode first sets up one PMP region to protect itself by setting the PMP configuration to be non-readable, non-writable, and non-executable, as shown in  \verb|pmp0cfg| in Fig.~\ref{fig:pmp-deployment}(a). 
Then, to allow the \sumode to access its own memory, \mmode sets up a second PMP region for OS and user-space, as per \verb|pmp1cfg| in Fig.~\ref{fig:pmp-deployment}(a). 
With this PMP configuration, any access by \sumode to SM will cause the \verb|pmp0cfg| check to fail, thus stopping accesses to \mmode. 
On the other hand, when the \sumode accesses region 1 (i.e., its own memory), \verb|pmp0cfg| does not apply; instead, \verb|pmp1cfg| applies and allows the access. 
In other words, by setting up regions and assigning permissions using priority-based PMP configurations, one can isolate the \mmode from \sumode on RISC-V.

Region configurations can overlap; in this case the hardware verifies an accessed memory address against the first matching configuration with the highest priority. 
Further, one can create multiple isolated regions within the \sumode by using multiple PMP configurations. 
Existing TEEs, such as Keystone, use this to create exclusive PMP regions for each enclave by flipping the access permissions between enclave and OS regions during context switch.
This is required to establish memory isolation between OS and enclaves, as PMP only differentiates between \mmode and \sumode, but not between different entities within \sumode.
At time $t_1$, before executing the OS, the SM grants S/U-mode access to the OS and denies access to the enclave.
Later at time $t_2$, before executing the enclave, the SM flips the permissions, allowing access to the enclave region and denying access to the OS.

\paragraph{Problem 1: Default access permissions of \mmode.}
PMP-based separation is unidirectional, i.e., only \mmode software is protected from access by unprivileged software but not vice versa. 
This leaves a large attack surface. 
We consider the following example where secure enclaves are deployed on the system. PMP enforces inter-enclave isolation, but the \mmode software is buggy. 
A malicious enclave exploits the bug (e.g., via ecalls) to get arbitrary code execution in \mmode (e.g., ROP chain). 
Since the code executes in \mmode, it can corrupt \mmode memory. More importantly, the enclave can trick the \mmode into 
accessing any \sumode memory (e.g., belonging to another enclave). 
Such an exploit could eventually lead to leaked or compromised data of a second enclave. 
If PMP-enforcement blocked \mmode's access to \sumode, the malicious enclave limits to \mmode.

\paragraph{Problem 2: \mmode PMP configurations are permanent.}
One way to address the above outlined problem is to  enforce PMP configurations on \mmode as well. In the PMP specification, this is indeed feasible, by setting an additional bit in the \verb|pmp0cfg| register. 
There are two undesirable repercussions of this: (a) once set, the PMP configurations that are applied to \mmode cannot be modified until reboot; (b) the configurations apply to all modes. 
Put together, this makes it impossible to change PMP configurations when switching enclaves (e.g., make only region 1 accessible at time $t_1$ and only region 2 at time $t_2$).
In the context of our example, when the enclave tricks the \mmode into accessing enclave 2's memory, the hardware will deny it. But enclave 2 itself can never access its own memory anymore. 
Thus, while this mechanism effectively limits \mmode's ability to access certain memory areas, it is not suitable for enclaves because they require frequent PMP configuration changes (e.g., switching execution between enclaves). 

\paragraph{Enhanced PMP (ePMP).}
It is a recently ratified extension to PMP~\cite{epmp} that 
applies each set of PMP configurations specifically to \mmode or \sumode.
Fig.~\ref{fig:pmp-deployment}(b) shows that ePMP allows setting up the memory regions where all entities have exclusive access to their own memory regions.
Further, ePMP allows dynamic changes to \mmode's PMP configuration. 
ePMP allows the flexibility needed to achieve \codename goals, but is not ideal. 
Specifically, it has three shortcomings: 
(a) any \mmode code can maliciously change regions and \pmpcfgs, leaving a large attack surface via buggy implementation;
(b) when applied, a PMP configuration is enforced for any code executing in \mmode, without fine-grained isolation within the mode; 
(c) even if fine-grained isolation is possible, transitioning between two partitions requires care of ePMP configuration to ensure isolation.

\section{\codename Compartments \& Interfaces}
\label{sct:compartments}

\codename introduces 2 compartments in \mmode: 
\textit{P} and \textit{F} isolate the SM and firmware from each other.
\codename enforces the 4 security invariants in Section~\ref{sect:problem_statement} to effectively prevent F from maliciously tampering with SM or enclaves.
However, housing these two compartments in \mmode requires careful consideration of the compartment memory layout, interfaces, transitions and permissions.
Therefore, \codename places each compartment in separate physical memory and ensures that F can never access any other memory outside its own compartment, satisfying \invmt.
For inter-compartment and inter-mode communication, \codename provides a set of secure register-based interfaces that P always checks and controls, thus satisfying \invenex.
To transition between compartments, \codename provides a secure mechanism that never unlocks both P's and F's memory spaces simultaneously, thus satisfying \invintf.
Finally, \codename introduces a binary scanner. During boot it ensures that the firmware is free from any malicious code that can change the isolation configuration, thus satisfying \invepmp.

\begin{figure}
    \centering
    \hspace*{-0.5cm} 
    \includegraphics[scale = 0.75]{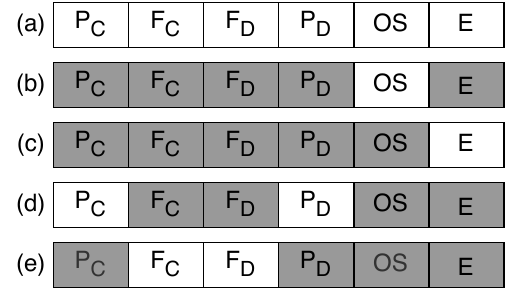}
    \caption{\codename access permissions when executing core. \textbf{P}=\compi, \textbf{F}=\compii, \textbf{OS}=Host OS, \textbf{E}=Enclave; C=code region, D=data region; \textbf{white}=access allowed, \textbf{grey}=access denied. (a)-(c): Same memory view as in \codename and legacy deployment: (a): Core operates in \mmode; (b): Core executes Host-OS in \sumode; (c): Core executes enclave in \sumode. (d)-(e): Memory views added by \codename: (d): Core operates in \mmode and the \primary executes; (e): Core operates in \mmode and the \secondary executes.}
    \label{fig:pmp-views}
    \hypertarget{figt:pmp-views}{}
\end{figure}

\paragraph{Memory Views.}
\codename starts with a memory layout such that 
code and data of the \compi are in separate physical memory regions without any overlap. Code pages are configured r-x, and data pages are configured rw-. 
Similarly, code and data for the \compii are separated in non-overlapping physical pages with the appropriate permissions. 
More importantly, \codename ensures that the \primary and \secondary do not have any code and data pages overlapping each other. 
\codename achieves these layout requirements by changing the bootloader and layout of the compiled \sm and firmware binaries. 

During execution, \codename ensures an isolated memory view, as shown in Fig.~\ref{fig:pmp-views}. 
First, it uses default \epmp features such that when P and F are executing, they cannot access S/U- and enclave memory.
Next, \codename confines P and F within the \mmode such that each can only access their own code and data. 
It again uses \epmp to achieve such intra \mmode isolation but in an atypical fashion.
For example, when the execution is transitioning from P to F, \codename disables access to P and enables access to F. 
More importantly, according to \invepmp, \codename only allows P to execute instructions related to PMP reconfiguration.
\codename also preserves Keystone's isolation model, i.e., \sumode cannot access \mmode and enclave memory. 
Lastly, for every interface that involves mode or compartment change, \codename changes the PMP configuration to reflect the correct memory view (Tab.~\ref{tab:interfaces} and Fig.~\ref{fig:pmp-views}).

\begin{table*}[]
\centering
\setlength{\tabcolsep}{1.3pt}
\caption{Transitions between compartments, OS and enclaves, that are affected by \codename}\label{tab:interfaces}
\resizebox{0.75\textwidth}{!}{%
\begin{tabular}{@{}lllllllllllllllllllllllll@{}}
\toprule
\textbf{Source} & \textbf{Destination} & \textbf{Event}   & \multicolumn{9}{l}{\textbf{Original Transitions}}      & \multicolumn{13}{l}{\textbf{\codename Transitions}}                            \\ \midrule
OS              & FW                   & ecall            & OS    & $\rightarrow$ & FW & $\rightarrow$ & OS   &     &    &     &       & OS    & $\rightarrow$ & P & $\rightarrow$ & F    & $\rightarrow$ & P & $\rightarrow$ & OS    &     &    &     &     \\

OS              & SM              & Enclave create/delete      & OS    & $\rightarrow$ & SM & $\rightarrow$ & OS &     &    &     &       & OS    & $\rightarrow$ & P & $\rightarrow$ & OS &     &   &     &       &     &    &     &     \\

OS              & Enclave              & Enclave enter      & OS    & $\rightarrow$ & SM & $\rightarrow$ & Encl. &     &    &     &       & OS    & $\rightarrow$ & P & $\rightarrow$ & Encl &     &   &     &       &     &    &     &     \\

Enclave         & OS                   & Enclave exit       & Encl. & $\rightarrow$ & SM & $\rightarrow$ & OS   &     &    &     &       & Encl. & $\rightarrow$ & P & $\rightarrow$ & OS   &     &   &     &       &     &    &     &     \\
Enclave         & OS                   & ocall            & Encl. & $\rightarrow$ & SM & $\rightarrow$ & OS   & $\rightarrow$ & SM & $\rightarrow$ & Encl. \ \ \ & Encl. & $\rightarrow$ & P & $\rightarrow$ & OS   & $\rightarrow$ & P & $\rightarrow$ & Encl. &     &    &     &     \\

Enclave         & FW                   & ecall            & Encl. & $\rightarrow$ & FW & $\rightarrow$ & Encl.   &  &  &  &   & Encl. & $\rightarrow$ & P & $\rightarrow$ & F & $\rightarrow$ & P & $\rightarrow$ & Encl. &     &    &     &     \\

OS              & -                    & Any M-Trap       & OS    & $\rightarrow$ & FW & $\rightarrow$ & OS   &     &    &     &       & OS    & $\rightarrow$ & P & $\rightarrow$ & F    & $\rightarrow$ & P & $\rightarrow$ & OS    &     &    &     &     \\
Leg. App \ \ \  & -                    & Timer M-Trap     & App   & $\rightarrow$ & FW & $\rightarrow$ & App  & $\rightarrow$ & OS & $\rightarrow$ & App   & App   & $\rightarrow$ & P & $\rightarrow$ & F    & $\rightarrow$ & P & $\rightarrow$ & App   & $\rightarrow$ & OS & $\rightarrow$ & App \\
Leg. App        & -                    & Non-Timer M-Trap & App   & $\rightarrow$ & FW & $\rightarrow$ & App  &     &    &     &       & App   & $\rightarrow$ & P & $\rightarrow$ & F    & $\rightarrow$ & P & $\rightarrow$ & App   &     &    &     &     \\
Enclave         & -                    & Timer M-Trap     & Encl.  & $\rightarrow$ & SM & $\rightarrow$ & OS   &     &    &     &       & Encl.  & $\rightarrow$ & P & $\rightarrow$ & OS   &     &   &     &       &     &    &     &     \\
Enclave         & -                    & Non-Timer M-Trap \ \ \ & Encl.     & $\rightarrow$ &  FW  & $\rightarrow$ & Encl.     &     &    &     &       & Encl.      & $\rightarrow$ & P  & $\rightarrow$ & F     & $\rightarrow$    & P  & $\rightarrow$    & Encl.      &     &    &     &     \\ \bottomrule
\end{tabular}%
}
\end{table*}

\begin{table}[]
\centering
\setlength{\tabcolsep}{1.3pt}
\caption{Unaffected Transitions}
\resizebox{0.85\columnwidth}{!}{%
\begin{tabular}{@{}llllllll@{}}
\toprule
\textbf{Source} & \textbf{Dest.} & \textbf{Event} & \multicolumn{5}{l}{\textbf{Unaffected Transitions}} \\ \midrule
App             & OS                   & ecall/syscall  & App          & $\rightarrow$    & OS    & $\rightarrow$   & App         \\
App             & OS                   & S-Trap         & App          & $\rightarrow$    & OS    & $\rightarrow$   & App         \\
Encl. App  \ \ \  & Runtime              & ecall/syscall \ \ \  & Encl. App    & $\rightarrow$    & RT    & $\rightarrow$   & Encl. App   \\
Encl. App       & Runtime              & S-Trap         & Encl. App    & $\rightarrow$    & RT    & $\rightarrow$   & Encl. App   \\ \bottomrule
\end{tabular}%
}
\end{table}

\begin{table}[]
\centering
\setlength{\tabcolsep}{4.5pt}
\caption{New transitions, introduced by \codename}
\resizebox{0.85\columnwidth}{!}{%
\begin{tabular}{@{}llllllll@{}}
\toprule
\textbf{Source} & \textbf{Dest.} & \textbf{Event} & \multicolumn{5}{l}{\textbf{New Transitions}}     \\ \midrule
P         & F            & Enter F              & P   & $\rightarrow$ & F $\rightarrow$   P        &  &    \\
F  \ \ \     & P              & Exit F               & F & $\rightarrow$ & P             &     &           \\
P         & -                    & M-Trap         & \multicolumn{5}{l}{impossible}                                 \\
F       &          -           & M-Trap   \ \ \       & F & $\rightarrow$ & F (trap handler) & $\rightarrow$ & F \\ \bottomrule
\end{tabular}%
}
\end{table}

\paragraph{Interfaces.}
Tab.~\ref{tab:interfaces} summarizes all the interactions between the components pertaining \codename. 
As per \invepmp and \invintf, any execution flow that incurs a memory view change has to transition through P.
This allows \codename to ensure that P makes the required \pmp updates to change the memory view without leaving any window of attack. 
For example, consider a scenario where an enclave wants to execute a syscall in the host OS. 
In Keystone, this involves transitions from enclave to \sm to OS. 
Since \codename executes \sm in P, a syscall interface already satisfies our requirement and is thus unaffected. 
Next, consider the case where the OS wants to invoke a function in the firmware.
In RISC-V, the OS simply makes an ecall to \mmode where the firmware services the OS request. 
In \codename, the firmware functions reside in the \compii. 
If the OS directly jumps to the entry point of F, the execution will fail because the firmware pages are inaccessible.
To address such cases, \codename enforces isolation during such transitions by first transitioning through P, satisfying \invenex. This way, for our ecall, P can lock the OS memory and unlock the firmware. 
\codename remedies all such flows as summarized in Tab.~\ref{tab:interfaces}.

\paragraph{Transitions between S/U and F.}
\codename has to enforce isolation between S/U- and \mmode. 
We observe that as per \epmp specification if the \pmp region is set as accessible for a particular mode (M or S/U), the other mode cannot access it. 
\codename leverages this to its advantage. 
When the execution is in \sumode, \codename sets the PMP configuration such that the region covers the entire memory of the currently executing unit (either OS or enclave). 
Similarly, \codename assigns \pmp regions that cover the entire memory of P's code and data. 
This way, when the \sumode switches to or from P, the \epmp enforcement automatically ensures mode-exclusive access. 
On RISC-V, the \sumode can execute an ecall to transition to \mmode. 
The CPU traps on the ecall instruction and uses the trap vector (stored in the \mtvec CSR) to locate and execute the handler that corresponds to the \mmode. 
By \invenex and \invintf, \codename enforces a fixed entry point into P by setting the trap vector correctly so an attacker cannot tamper it. 
As changing the trap vector requires \mmode privileges, \codename ensures that only P and not F can change it. 

\paragraph{Transitions between OS and enclave.}
Keystone execution flow for transitions between enclaves and OS is mediated via the \sm. Since \codename just moves the \sm into P, we do not have to change \pmp switching or transition flow. 

\paragraph{Transitions between F and P.}
So far we have looked at inter-mode transitions (S/U vs. M) 
or intra-mode transitions that are forced to incur a mode transition (OS to enclave via \mmode). 
However, the critical change brought about by \codename is the creation of intra-mode isolation between F and P. 
Our interface, respecting \invintf, only allows P to invoke F, and that only for certain services such that F performs the operations and simply returns to P. This choice reduces several attack vectors (e.g., multiple entry points to F from OS/enclave/P, or arbitrary calls from F to P).
Nonetheless, \codename still has to facilitate a call from P into F and return from F to P. 
This has to be done with care for two reasons: (a) P and F execute in \mmode, and both have the privileges to change \pmp configuration; (b) during transition between compartments, memory and interrupt isolation has to be enforced without leaving a window for attacks (e.g., TOCTOU, lack of atomicity). 

\paragraph{Bootstrapping.}
On traditional RISC-V systems, the first stage bootloader (FSBL) loads and starts the firmware, which in turn starts the OS. 
\codename modifies this bootprocess, particularly the FSBL to enforce \invepmp and \invmt.
In \codename firmware and SM consist of two distinct binaries. The FSBL loads both of them into memory.
But before starting any software in \mmode, the FSBL scans the firmware's code blob to ensure that it does not contain any malicious gadgets that can reconfigure PMP-related registers or the trap vector.
Since RISC-V instructions are 2\,B aligned, \codename scanner checks the blob for suspicious opcode patterns in 2\,B intervals.
If the scan does not detect any such opcodes, the FSBL starts the SM binary.\footnote{This one-time scan at boot is sufficient because F cannot write to its own code memory or execute data memory during runtime.
}
The SM's initialization creates the P and F compartments by configuring the PMP registers.
This creates and enforces the isolation between SM and firmware.
After P is initialized, \codename transitions to F to start the initialization of the firmware.
When F finishes, \codename transitions back to P, which then starts the OS.

\section{Inter-compartment Transitions}
\label{ssec:transitions}

The transitioning process between the \primary and the \secondary requires differentiating between the cases of switching from P to F and from F to P to ensure a correct switching routine as required by \invepmp and \invmt.

\subsection{PMP to Firmware Compartment}
\label{sec:p2s}

Switching from P to F entails four steps: removing access to P’s memory regions, allowing access to F’s memory regions, jumping to F’s entry point, and setting the trap handler to F’s view. Despite the universal trust in P, these steps require careful consideration to prevent malicious influence from S/U-mode or F. 

\paragraph{Execution in P.} 
P is executed either because of transition from \sumode or because it returned from F. 
When entering into P from \sumode, \codename must ensure that for every mode the permissions for \sumode memory regions are disabled.
This \pmp configuration change happens directly after entering P from \sumode.
Thus, when P is executing, it can only access its own memory (code and data pages).
Additionally, as per RISC-V ISA~\cite{riscvSpecs}, the CPU automatically disables interrupts when entering \mmode.
This means that entering \mmode from \sumode will automatically mask all interrupts.
The \mmode has to explicitly re-enable them by writing to a specific CSR.
In \codename, we never enable interrupts while executing in P.
Thus, P never incurs traps: 
(a) exceptions such as divide-by-zero never arise because of careful programming; 
(b) interrupts such as timers, even if triggered, are masked. 
Lastly, P has its own view of the trap vector---it has to be set such that the ecall entry point is fixed. 
Put together, \codename ensures that P has memory and trap isolation as specified by \invmt.

\paragraph{Entering F.} 
Next, we consider the case where P invokes a function in F, which requires a transition from P to F, which entails three steps: 
Reconfigure PMP to deny access to P's memory regions and allow access to F's memory regions, jump to F's entry point, and set the trap handler to F's view. 
The order of these steps is crucial for several reasons.
\codename does not allow F to change the trap handler to preserve security, as we will explain later in Section~\ref{sec:s2p}, so P has to change it before entering F. 
If P removes access permission to its own memory regions, it can no longer perform \pmp configuration changes to allow access to F's memory regions. So it has to allow access to F's regions before or together with removing access to its own regions. 
Similarly, P cannot perform the jump to F's entry point before or after removing access to its own regions. 
Next, we explain how \codename addresses these challenges securely. 

\paragraph{Reconfiguring ePMP.} 
The crucial step of switching from P to F is the re-configuration of \epmp.
Essentially, \codename has to perform a configuration transition as visualized in Figs.~\hyperlink{figt:pmp-views}{\ref{fig:pmp-views}d}--\hyperlink{figt:pmp-views}{\ref{fig:pmp-views}e}, thus stopping the execution in P and starting the execution in F.
To achieve this, P needs to prepare the \pmp configuration registers such that the isolation hardware enforces the combination mentioned above.
P first writes required values into a general-purpose register, which it then moves into the \pmp configuration register.
Note that this reconfiguration can be performed in a single step
to disable access to P's regions and enable access to F's memory regions without an intermediate step where both P and F are accessible. 
If \codename performs this step with two instructions, e.g., first allowing access to F's memory regions and afterwards removing access to P's memory regions, we would have a time window in which both regions are accessible at the same time.
If now, for whatever reason, an exception occurs, the CPU would start executing F's trap handler since we have already set the reference to it in the trap vector in the previous step.
Since both P's and F's memory regions are accessible now, and F is executing, it could gain complete control over \mmode, which would contradict \invmt.
However, as explained above, it is impossible that \codename will trigger a trap (exception or interrupt) during these two operations. 
Hence, in this case, changing the PMP configurations does not necessarily need to be performed atomically.

\begin{figure}
    \centering
    \hspace*{-0.45cm}
    \includegraphics[width=1.06\columnwidth]{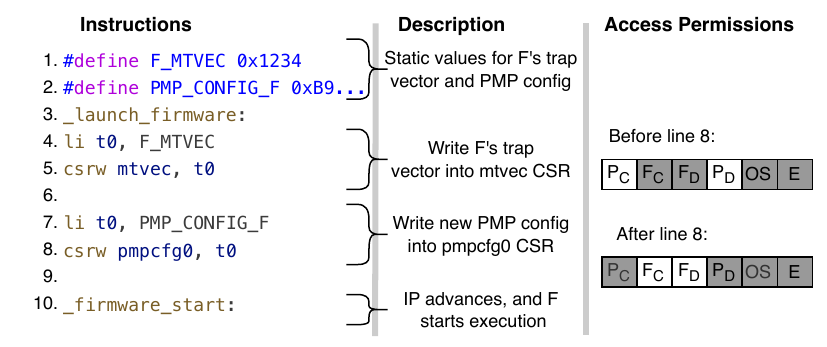}
    \caption{\textbf{Transition P to F}. P updates trap vector to F's view and then reconfigures access permissions for code and data regions of P and F atomically using static configuration value.}
    \label{lst:primary_exit}
\end{figure}

\paragraph{Jumping to the Firmware Compartment.}
The last crucial step in switching from P to F is to set the instruction pointer (IP) to the start of F's code region.
However, the question remains if setting the IP should be performed before or after changing the \pmp configuration.
The main issue is that since we set F's and revoke P's access permissions, we cannot execute code any further in P after this.
Therefore, we make use of the CPU's notion of advancing the instruction pointer automatically after executing an instruction.
We place the instruction that changes the PMP configuration into the last few bytes of P's code region.
Executing these instructions directly at the border has the effect that we directly advance the IP from P's code region into F's code region while simultaneously re-configuring the memory access permissions.
Fig.~\ref{lst:primary_exit} shows an assembly pseudocode example of performing the overall switching from P to S. 

\subsection{Firmware to PMP Compartment}
\label{sec:s2p}

Switching back from F to P entails four steps: 
remove access to F's memory regions, allow access to P's memory regions, jump to P's entry point, set trap handler to P's view.
It requires careful consideration when performing these security-sensitive tasks since the attacker controls F.
The attacker cannot achieve arbitrary code injection (\codename enforces DEP using \epmp).
But it can perform arbitrary code execution, by exploiting memory vulnerabilities to achieve ROP.

\begin{figure}
    \centering
    \hspace*{-0.45cm}
    \includegraphics[width=1.06\columnwidth]{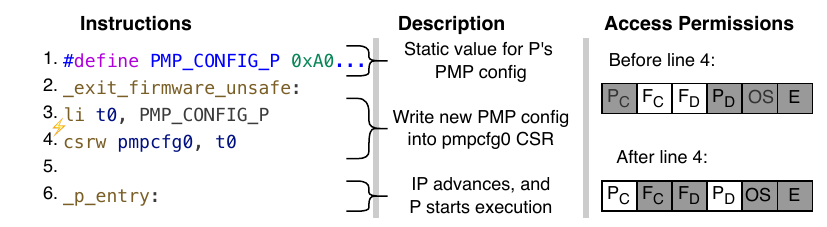}
    \caption{\textbf{Unsafe Transition from F to P}. F reconfigures ePMP with static value to deny access to F's regions and grant access to P's regions. ROP can exploit this transition as a gadget by jumping between lines 3-4. }
    \label{lst:secondary_unsafe_exit}
\end{figure}

Let us first consider the step where F removes access permissions to its own memory regions and allows access to P's regions.
Fig.~\ref{lst:secondary_unsafe_exit} shows a pseudocode for performing this operation.
First, \codename does not allow F to execute any instructions that change \pmp configurations. 
Second, even if we selectively allow F to execute the particular sequence of \pmp-related instructions in Fig.~\ref{lst:secondary_unsafe_exit}, we cannot trust F to faithfully prepare the configuration in a general-purpose register and then move it into the \pmp configuration register.
F can abuse this to write arbitrary values into the PMP configuration registers. 
Therefore, Fig.~\ref{lst:secondary_unsafe_exit} provides a very expressive gadget that an attacker could use to change \pmp configurations arbitrarily. 
This raises the dilemma that F needs to perform \pmp reconfigurations, even though it is not trusted to do it.

\paragraph{Reconfiguration of PMP using a SallyPort}
The main challenge in solving this dilemma is to achieve the same effect that doing a \pmp configuration change has, but with instructions that cannot be misused as a gadget.
We address this by introducing the notion of a {\em SallyPort} region.\footnote{Sally port can be entered but not exited through the same door.}
First, we construct a way to revoke F's access with a single write into the PMP configuration register that has a static value. This way, even if the attacker uses this mechanism as a gadget, it will always end up getting access to its own memory regions removed.
We refer to this as a {\em SallyPort-Entry}, shortened to SPEntry; F can use it, but it only revokes access to regions and never explicitly adds new permissions.

\begin{figure}
    \centering
    \hspace*{-0.45cm}
    \includegraphics[width=1.047\columnwidth]{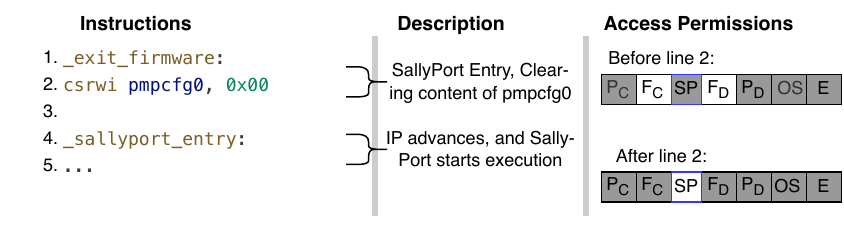}
    \caption{\textbf{Transition from F to SP using SPEntry}. F clears \texttt{pmpcfg0}, removing access to F's memory regions and entry that denies access to SP. Now the core can access SP.}
    \label{lst:secondary_exit}
\end{figure}

Next, \codename has to enable access permissions of P's memory regions and jump to P's entry point. But the \pmp-related instruction to enable it cannot be placed in F; it has to be executed after F is locked. 
We again use the insight of instruction pointer increment. 
We slightly modify the SPEntry such that it atomically locks F safely and unlocks P.
It achieves this atomicity by putting the instructions that change \pmp configurations at the end of F's code region and placing a part of P's code region immediately after the end of F (Fig.~\ref{lst:secondary_exit}). 
This creates a layout where F is sandwiched between P's code pages. 
Specifically, we refer to the part of P's code region that resides immediately after F's code region as the SallyPort (SP).
At this point, \codename can perform the remaining tasks of restoring P's trap vector, jumping to the entry point of P, and continuing its execution. 

We now detail the construction of such a SPEntry.
\codename uses a particular insight about setting PMP configurations to achieve atomic locking of F and unlocking of P.
We observe that PMP bundles the permission setup for eight consecutive regions in one \verb|pmpcfg| register.
This means that memory regions 0-7 (with high priority) are configured using \verb|pmpcfg0|, while regions 8-15 (with lower priority) are configured with \verb|pmpcfg2|.\footnote{\texttt{pmpcfg1} is invalid / does not exist on 64-bit RISC-V CPUs} 
\codename can ensure that permissions for P and F are exclusively configured with the \verb|pmpcfg0| register (i.e., all code and data regions for P and F are always protected by \pmp entries between 0 and 7).
In such a case, if F writes zeroes to \verb|pmpcfg0|, it will have the effect that regions 0-7 are essentially non-existent, i.e., not configured.
Thus, any access to memory previously covered by regions configured with \verb|pmpcfg0| is denied if no configurations exist for them in \verb|pmpcfg2|.
We can clear \verb|pmpcfg0| using the immediate-CSR-write instruction, which does not take an argument from a register but rather from one that is embedded in the instruction itself.
Concretely, for our case, this is the instruction \verb|csrwi pmpcfg0, 0|.
This instruction cannot be used as a modifiable gadget, as it will always perform the same operation since no general-purpose register can be specified.
Further, after executing the instruction that writes into the \verb|pmpcfg| register, the CPU will increment the IP, say to an address \textit{M}. 
If \textit{M} falls within the range covered by \verb|pmpaddr0-7|, the instruction execution will cause an access fault due to \pmp violation. 
On the other hand, if \textit{M} falls within the range covered by \verb|pmpaddr8-15|, the CPU will enforce the PMP configuration as defined in \verb|pmpcfg2|. 
If memory access permissions for F are completely covered with \verb|pmpcfg0| and at least some parts of P's memory access permissions are covered with \verb|pmpcfg2|, the above gadget will achieve our desired goal.
To this end, we place the SPEntry as the last instruction in F at the end of the page followed immediately by SP.
Fig.~\ref{lst:secondary_exit} showcases a full transition in pseudocode from F to SP.

\begin{figure}
    \centering
    \hspace*{-0.41cm}
    \includegraphics[width=1.04\columnwidth]{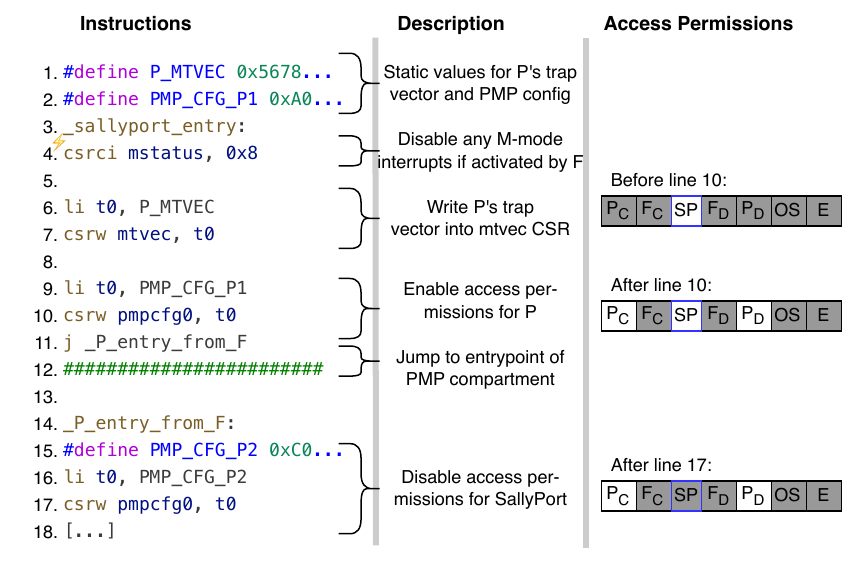}
    \caption{\textbf{Transition from SP to P}. The SP reconfigures the trap vector for P and restores access permissions to P's regions. Afterwards, SP jumps to P, which in turn removes access permissions for SP's memory region. An interrupt set by F could be triggered between lines 4-5.}
    \label{lst:jumppad_to_p}
\end{figure}

\paragraph{Switching from the SP to P.}
F uses the SPEntry to switch into SP, which then unlocks P and restarts the execution of P.
\codename performs several steps in the SP.
First, it disables any interrupts that might be active in \mmode.
Then, it resets the trap vector to P's handler.
Eventually, it reinstalls required PMP configurations in \verb|pmpcfg0|. 
Finally, it jumps to a pre-specified entry point address of P.
Note that in this case, we are not using any tricks to advance the instruction pointer. 
Instead, we trust the SP to perform the PMP configuration changes faithfully and to normally jump to P's entry point.
Fig.~\ref{lst:jumppad_to_p} shows the transition from SP to P.

\paragraph{Why is F not allowed to change the trap vector?}
Recall that F and P need to have their own views of the trap vectors for isolation. 
One way to achieve this is to simply save and restore the trap vector between compartment switches. 
However, \codename chooses a design where:
At initialization, the trap vector is set to P's value of choice. 
Right before P transitions to F, it updates the trap vector to F's view that P deems safe---all handlers in the trap vector must land in F's code pages.
In other words, \codename never allows F to change the trap vector to the extent that it employs a binary checker to ensure that F is launched if and only if it has no such instructions. This measure is to stop the following attack. 

Recall the mechanism of switching from F to P, where we use two instructions in SP to a) set the \pmp configurations and b) disable any \mmode interrupts.
This opens up a window where the CPU can receive interrupts (e.g., a timer activated intentionally by F just before exiting).
At this point, the trap vector is still set to a value controlled by F, which could point to any arbitrary code available in SP. 
If the timer fires between the PMP-configuration and interrupt-disabling instructions, the CPU will execute the code pointed to by the trap vector, set by F, which will execute instructions in SP at a location of F's choice and with register states provided by F.
However, if the trap handler still points to the original location in F, as P has set it before transitioning, and an interrupt arrives, the CPU will only stall unrecoverably until reset and prevent this attack.
That is because the trap handler itself is located at an address that is now no longer accessible as the PMP configuration was just cleared. So the CPU is forced to endlessly trigger access fault exceptions.
This is why \codename's design never allows F to set the trap vector.
However, this does not prohibit F from handling interrupts and exceptions---P, who is trusted to set the trap handler sets it just before entering F (as shown in Section~\ref{sec:p2s}).

\paragraph{Binary Scanning.}
We observe that, on RISC-V, the only way to modify the trap vector (\mtvec) is to perform a CSR-write instruction for this register.
Therefore, \codename scans F during boot to detect if CSR-write instructions are being used in combination with \mtvec.
This scan is in addition to the one that checks if F performs any modifications to the \pmp configuration, with the only exception of the SPEntry.

\section{Security Analysis}
\label{sec:sec-analysis}

We argue the security of \codename from various attackers.

\subsection{Malicious S/U-Mode}
\codename's P ensures security of host OS and enclaves using ePMP, exactly as in Keystone~\cite{keystone}. 
The host OS or enclave runtime can try to attack the 
\mmode (both compartments) by directly tampering with ePMP registers. The CPU will disallow this because they do not have the privileges to access ePMP registers as per RISC-V specification (\invepmp).

S/U-mode can make ecalls to invoke P and  misconfigure ePMPs. Since we assume that P is bug-free and performs input sanitization, the S/U-mode cannot trick P into doing their bidding (e.g., change \verb|pmpcfg0|).
S/U-mode can also try to use F to divert execution 
into a different entity i.e., host-OS to enclave or vice versa, without P's supervision.
P prevents this attack as it strictly controls that only host OS can invoke F's functions.
Further, since P is the only entry and exit point of \mmode, it can ensure that F can never compromise the execution flow (\invenex and \invintf).

S-mode can try to invoke \compii directly, either by jumping to the physical memory address or changing the \mtvec register for re-routing ecalls.
However, directly jumping to F's code triggers an access fault exception: (i) the PMP configuration prevents any direct accesses from lower execution levels; (ii) changing the \mtvec register will trigger an illegal instruction exception.
Since both exceptions are diverted to P, it will deny any such attempts (\invenex).
S-mode can use F to launch attacks against P as we discuss next.

\subsection{Malicious Firmware Compartment}
\mmode is used by the \compi and \compii. 
Since P is considered trusted, we focus on the attacks that F can launch against P or enclaves.
First, F can attempt to access memory beyond its PMP region configuration (i.e., the memory of P, the host OS or enclaves).
The CPU triggers access fault exceptions to deny access.
Although these exceptions trigger \compii's exception handler, it cannot successfully recover since the handler itself does not contain any PMP-related instructions as confirmed from the binary scanning during boot  (\invepmp and \invmt).

F's attempts to orchestrate Iago attacks to trigger PMP reconfigurations in P are likewise ineffective. 
P does not initiate PMP reconfigurations based on F's return values, and transitions between P and F adhere to a rigid program flow without dynamic control flow changes, as enforced per \invintf.

F can generate malicious instructions that alter PMP configuration or collude with the OS to copy code blobs containing them.
F would attempt to write these instructions to its code or data sections and then start their execution.
However, both trying to write to the code section or executing data as code will trigger an access fault exception.
Similar to the previous case, the exception handler of F cannot recover from this exception since it does not contain any PMP-related instructions (\invepmp and \invmt).
Maliciously using code sections from the enclave or host OS to generate malicious instructions is also ineffective, as this memory is not accessible to F.
Any attempts to access leads to access fault exception that F cannot recover from.

F can attempt ROP-style attacks and execute malicious instructions as gadgets.
It will be unsuccessful because: a) RISC-V has a fixed-length instruction set of 2 or 4 Bytes. Any unaligned execution will cause an exception that F cannot handle for the same reasons mentioned above. b) \codename ensures using binary scanning during boot that no gadgets exist that the attacker could exploit to reconfigure PMP.

F can try to execute PMP instructions in the SP by exploiting the non-atomicity of interrupt masking as follows:
after switching into SP but before SP masks the interrupts, F schedules a timer interrupt that diverts execution back F's handler. 
F then uses the instructions in the now unlocked SP as a gadget to maliciously reconfigure PMP.
This attack is not successful, as F's trap handler is no longer accessible to the CPU after switching into SP (\invintf and \invmt).
So, the CPU will trigger an infinite loop of access fault exceptions from which the CPU cannot recover until a hard reset.

F can try \mtvec writes to modify the address of the exception handler triggered prior to invoking SP.
However, binary scanning during boot prevents the existence of any such instructions in F, and generating it faces the same challenges mentioned above in regard to the PMP-related instructions.

\section{Implementation}
\label{sec:impl}

Implementing \codename involves assessing both the hardware features available on the targeted platform and existing software components, including the bootloader and firmware.

\subsection{Components and Placement} 
\label{ssec:components}

We implement \codename on the HiFive Unmatched board. 
We extend the platform’s stage-1 bootloader (U-Boot SPL, provided by the manufacturer) with the binary scanner.
The \compi in the M-mode houses Keystone’s SM, extracted from the OpenSBI extension.
The \compii accommodates a modified OpenSBI version, featuring adjustments for interaction with the \compi.
We disable OpenSBI’s functions responsible for relocating itself since we assume that the memory areas are fixed and the software in M-mode is correctly placed in memory by trusted bootloaders in earlier stages.
To pass the measurement during boot, we remove any instructions related to modifications of the PMP entries or the trap vector since the \compi now performs these operations instead.
Data exchange between \compi and \compii solely happens using registers passed during invocation of the \compii (P to F), and after it yields (F to P).

\subsection{Platform Considerations}
\label{ssec:plat-considerations}

\codename's design requires ePMP extensions to enforce the co-isolation of the \compi and the \compii in \mmode. 
We tested our implementations on platforms that natively support ePMP: QEMU and NOEL-V~\cite{noelv}. 
Further, we modified Rocketchip to add ePMP support.
During our tests, we did not detect any memory access violations.

However, none of our ePMP-enabled platforms are suitable for evaluating \codename.
QEMU is not cycle accurate, its performance changes based on host OS.
Cores running on FPGA require significant engineering to port Linux and Keystone if they do not support it (e.g., NOEL-V). 
For cores that do support Keystone (e.g., Rocketchip), our ePMP implementation is not optimized and exhibits large variance even for baseline execution i.e., without \codename changes.

Ideally, we need a production board that supports ePMP.
However, at the time of writing, we could not obtain any board that supports ePMP.
Therefore, we evalaute \codename on a platform with only PMP support i.e., the HiFive Unmatched.
While the lack of ePMP prevents a direct implementation of \codename as per our original design, we adapted our approach to demonstrate the feasibility.
We modified the PMP configurations during context switches to exclude the L-bit, effectively activating PMP configuration enforcement only for S/U-mode. 
We do not expect any difference in performance in terms of executed cycles by using this board compared to one that employs ePMP (see Section~\ref{sec:eval-epmp}).

\paragraph{Number of ePMP entries, enclaves, and compartments.}
\codename creates two M-mode compartments: P and F.\footnote{See Section~\ref{disc:ncompartments} for extending \codename to multiple F compartments.}
This requires 6 PMP entries during execution: 1 for non-enclave S/U-mode (e.g., OS) and 5 for M-mode (2 for P's code and data, 2 for F's code and data, 1 for SallyPort). Fig.~\ref{lst:secondary_exit}--\ref{lst:jumppad_to_p} show this in detail.
Since \codename needs one of these entries to be the 9$^{th}$ entry, we need at least 9 entries on the platform.
Further, each enclave needs its own PMP entries.
For example, Keystone requires at least 2 PMP entries per enclave for storing enclave private and host shared data, respectively~\cite{keystone}, whereas Elasticlave needs 3 PMP entries per enclave~\cite{elasticlave}. 
\codename does not change this requirement per enclave.

\section{Evaluation}
\label{sec:eval}

\begin{table}[]
  \centering
  \caption{LoC Breakdown of M-Mode software for a standard Keystone deployment and for our \codename prototype. For Keystone, we removed unrelated device-specific libraries from Keystone SM for fair comparison.}
  \resizebox{0.75\columnwidth}{!}{%
  \begin{tabular}{lrr}
  \toprule
  \textbf{Component}                  & Keystone & \textbf{\codename} \\ \midrule
  OpenSBI                             & \textbf{23942}          & 23922                 \\
  Dorami Compat Stubs                 & -                       & 46                    \\ \arrayrulecolor{black!30}\midrule
  Keystone SM                         & \textbf{7777}           & \textbf{6777}                  \\
  OpenSBI Stubs                       & -                       & \textbf{2724}                  \\
  Dorami Enforcement                  & -                       & \textbf{1156}                  \\ \arrayrulecolor{black}\midrule
  \textbf{TCB}                        & \textbf{31751}          & \textbf{10657}        \\ \midrule
  Total                               & 31751                   & 34625                 \\ \bottomrule
  \end{tabular}
  }
  \label{tab:loc}
\end{table}

We show impact of using ePMP instead of PMP (Section~\ref{sec:eval-epmp}). 
and \codename's performance (Section~\ref{ssec:perf}).

\paragraph{TCB.} \codename in total consists of 34.6K\,LoC; this includes a) the \primary that houses the PMP reconfiguration functions and the SM for enclave functions and b) the \secondary compartment that houses OpenSBI.
Tab.~\ref{tab:loc} summarizes the lines of code of \codename compared to a typical system setup. 
We provide detailed information on the impact of the individual monitor components and LoC breakdown.

\paragraph{Bootstrapping the PMP compartment.}
A considerable part of the implementation overhead of the \primary is the code required for setting itself up and for basic LibC-like functions that are required to operate. We implemented an initialization routine similar to the one provided by OpenSBI with 343\,LoC (included in \codename Enforcement in Tab.~\ref{tab:loc}). We import various functions (OpenSBI stubs) for the SM to function correctly.

\paragraph{Calling Firmware Services from the Host-OS.} 
One of P's main tasks is to forward interrupts and exceptions from S/U-mode to F, as showcased in Tab.~\ref{tab:interfaces}. 
The switching process involves saving the context from S/U-mode, creating and loading a context for F and starting F. 
After F yields, P needs to save and evaluate the context it returns.
Based on this data, P modifies the previously saved context from S/U-mode and restores the context before switching to it. 
This functionality is part of the \codename Enforcement in Tab.~\ref{tab:loc} and consists of 813\,LoC: 474\,LoC written in C and 339\,LoC in assembly.

\paragraph{Handling Keystone Requests.} 
Adding the Keystone SM to the \primary allows it to service the OS with launching and managing secure enclaves. Including Keystone's SM into P required adding library functions from the original OpenSBI in the amount of 2724\,LoC.

\subsection{Impact of \epmp}
\label{sec:eval-epmp}

\begin{table}[]
  \centering
  \caption{Synthesis overview for Rocketchip and NOEL-V in different configurations. SI: Single-Issue, DI: Dual-Issue. All NOEL-V SI configurations synthesized with 39\,DSPs and 84.5\,BRAMs. All NOEL-V DI configurations synthesized with 39\,DSPs and 96.5\,BRAMs. All Rocketchip configurations synthesized with 36\,DSPs and 350\,BRAMs. The gray lines show the maximum capacity of the FPGA.}
  \label{tab:syn_impact}
  \resizebox{0.7\columnwidth}{!}{%
  \begin{tabular}{@{}lrrrr@{}}\toprule
  \multicolumn{1}{c}{\textbf{Setup}} &
    \multicolumn{1}{c}{\textbf{\begin{tabular}[c]{@{}c@{}}No.\\ PMP\end{tabular}}} &
    \multicolumn{1}{c}{\textbf{\begin{tabular}[c]{@{}c@{}}CLB\\ LUT\end{tabular}}} &
    \multicolumn{1}{c}{\textbf{\begin{tabular}[c]{@{}c@{}}CLB\\ FF\end{tabular}}} &
    \multicolumn{1}{c}{\textbf{\begin{tabular}[c]{@{}c@{}}F-Mux\\ {[}7+8{]}\end{tabular}}} \\ \hline
  \rowcolor[HTML]{EFEFEF} 
  \multicolumn{2}{l}{\cellcolor[HTML]{EFEFEF}VCU118}                               & 1182240 & 2364480 & 886680 \\ \hline
                                                                              & 0  & 170642  & 150288  & 4202   \\
                                                                              & 15 & 176307  & 151489  & 4228   \\
  \multirow{-3}{*}{\begin{tabular}[c]{@{}l@{}}ROCKET\\ SI, PMP\end{tabular}}  & \% & 3.31    & 0.79    & 0.62   \\ \hline
                                                                              & 15 & 176378  & 151497  & 4264   \\
  \multirow{-2}{*}{\begin{tabular}[c]{@{}l@{}}ROCKET\\ SI, ePMP\end{tabular}} & \% & 3.36    & 0.80    & 1.47   \\ \hline
  \rowcolor[HTML]{EFEFEF} 
  \multicolumn{2}{l}{\cellcolor[HTML]{EFEFEF}KCU105}                               & 242400  & 484800  & 181800 \\ \hline
                                                                              & 0  & 128907  & 73389   & 5053   \\
                                                                              & 15 & 131973  & 75890   & 5621   \\
  \multirow{-3}{*}{\begin{tabular}[c]{@{}l@{}}NOEL-V\\ SI, ePMP\end{tabular}} & \% & 2.37    & 3.41    & 11.24  \\ \hline
                                                                              & 0  & 144233  & 77179   & 4165   \\
                                                                              & 15 & 150841  & 79943   & 5506   \\
  \multirow{-3}{*}{\begin{tabular}[c]{@{}l@{}}NOEL-V\\ DI, ePMP\end{tabular}} & \% & 4.58    & 3.58    & 32.19  \\ \bottomrule
  \end{tabular}%
  }
  \end{table}

We synthesize the custom Rocketchip CPU for a Xilinx VCU118 FPGA using FireSim and the NOEL-V CPU for a Xilinx KCU105 FPGA using Gaisler GRlib, both with a target frequency of 100\,MHz. 
Both CPUs feature two RV64GC cores with \epmp support for up to 15 entries each.\footnote{Our NOEL-V synthesis for 16 PMPs failed so we used 15 for all instead.}
We synthesize one PMP-only and three ePMP-enabled setups summarized in Tab.~\ref{tab:syn_impact}, varying from 0 to 15 \pmp entries.
We report the impact of enabling ePMP support on area and PMP/ePMP reconfiguration on four setups. 
We test our setups using the CLI Linux distributions provided by FireSim and GRlib that are compatible with Rocketchip and NOEL-V. Rocketchip operates with Linux v.\,6.2 and OpenSBI v.\,1.2, and NOEL-V with Linux v.\,6.8 and OpenSBI v.\,1.4.

\paragraph{Area Impact.}
Our first goal is to measure the impact of introducing PMP to a core.
We use Rocketchip to evaluate this by configuring it with 0 and 15 PMP entries.\footnote{We outline security
considerations for integrating ePMP in Section~\ref{disc:microarch}.}
As shown in Tab.~\ref{tab:syn_impact} (rows 4 vs 5), there is a large increase in LUT and F-Mux usage, purely due to PMP logic. 
Next, we measure the impact of moving from PMP to ePMP. Our Rocketchip measurements with our ePMP implementation show that this incurs relatively less hardware overheads (additionally 0.04\,\% LUTs and 0.85\,\% F-Mux).
This shows that while introducing PMP logic does have high impact, extending that logic to ePMP support is not invasive. 
Lastly, we measure the impact of ePMP for different pipelines by 
configuring NOEL-V in single and dual-issue modes.
When we increase the entries from 0 to 15 we see that dual-issue NOEL-V requires nearly twice the amount of additional LUTs and thrice the amount of additional F-Muxes compared to single-issue.
From this, we conclude that the impact of ePMP can increase with a more complex pipeline. 
We note that for all our experiments, for a particular configuration of core, the number of PMP and ePMP entries do not have an impact on DSPs and BRAMs.

\paragraph{Execution Overheads.}
Our goal is to measure the effect of frequently changing ePMP configurations and how it varies for different core implementation. 
To evaluate this, we used three setups:
Rocketchip with our ePMP implementation, NOEL-V with in-built ePMP implementation in single and dual-issue mode. 
We used the AES binary from RV8 benchmark~\cite{rv8-bench} and performed a configuration change for every context switch incurred for scheduling. Specifically, we invoked an $\tt{ecall}$ which performs a write to the \verb|pmpcfg0| register to incur a configuration change. 
For each setup, we ran a baseline which performs the ecall but does not change the configuration (i.e., no writes to \verb|pmpcfg0|).
We report an overhead of $5.58\%$ for RocketChip, whereas it is only $1.87\%$ for NOEL-V in single-issue mode. 
This shows that the impact of ePMP varies based on the implementation. 
One possible explanation why Rocketchip introduces higher overhead is that our implementation of ePMP has not been optimized whereas the NOEL-V's implementation is production-level.
Next, we report an overhead of $0.54\%$ for NOEL-V in dual-issue mode. 
Thus, adding ePMP checks does not slow down optimized pipelines as much as in the case of single-issue.

\subsection{Performance}
\label{ssec:perf}

We measure the lifecycle cost for booting the platform, host OS, and enclave operations. 
Then we benchmark stress tests for CPU with RV8 and I/O with FSCQ. 
Lastly, we measure the performance impact on real-world applications: webservers with darkhttpd and in-memory databases with SQLite.

\paragraph{Hardware \& Software Setup.}
All our evaluation numbers are presented based on 10 runs measured on the
HiFive Unmatched board four SiFive U74 RV64GC cores~\cite{u74}.
The U74 cores operate at 1.2 GHz with 32KB instruction and data caches, with 8 PMP entries per core.
While \codename needs at least 9 entries which would incur two writes, our board only needs one write to update the PMP entries. 
To capture the cost of two writes, we add an additional write instruction in our code that performs the same write twice while ensuring that the compiler does not optimize it out.
For software setup, we use the SiFive Freedom-U-SDK(2022.10)~\cite{freedomusdk} for building a CLI Linux distribution (with Kernel v.\,5.19) as the OS and OpenSBI (v.\,1.3) as the firmware.
We build two setups for the measurements: baseline and \codename.
The baseline setup consists of a standard Keystone deployment. 
At the time of writing, Keystone does not provide a functional/stable base configuration for Unmatched board. Therefore, we manually add Keystone's SM to the OpenSBI version of the SDK as an extension. We modify the Kernel configuration to support contiguous memory allocations as Keystone requires. Additionally, we add Keystone's kernel module using a new yocto recipe.
\codename setup is a standard deployment of \codename. We place the software that handles the PMP configurations and enclave management in the \primary as described in Section~\ref{sct:compartments}. In the \secondary, we place the modified OpenSBI, which performs no PMP configurations except for using the SPEntry, as described in Section~\ref{sec:s2p}.
We use the same OS as in the baseline.

\paragraph{Measuring Lifecycle Costs.} 
Since \codename effectively acts as an additional layer in the firmware and lies on the critical path for handling M-mode interrupts or exceptions, we expect it to directly impact the overall system performance.
Tab.~\ref{tab:overhead_minibench}~(left) summarizes the overheads of \codename for life-cycle operations for the platform, host, and enclaves.

\paragraph{Platform Initialization Cost.} 
During firmware boot, \codename adds a minimal overhead of 20.81,\% to set up its data structures and to set up static memory access configurations of the \primary, the \secondary, and host-OS. This overhead is a one-time cost during system boot-up and does not repeat until a restart of the system. 

\paragraph{Host-OS Boot Cost.} 
While booting the host-OS, the kernel performs several calls into the firmware to get information about the system or reconfigure settings in the firmware. 
During kernel boot we measured the overhead for executing the base platform handler in OpenSBI that returns information about the platform and OpenSBI itself.
The first entry in Tab.~\ref{tab:interfaces} shows the typical transitions involved in requesting such services in the firmware.
Although we observed a slowdown of 4-10x for these ecalls (Tab.~\ref{tab:overhead_minibench}~left, rows\,5-6) it did not have a major impact on the overall boot process of the OS.
The reason is that during general runtime, the \mmode is mainly only invoked because of timer interrupts that the OS requests, with a frequency of 251 interrupts per second.
Rows 6--7 in Tab.~\ref{tab:interfaces} show which transitions would be involved when an M-mode interrupt triggers while executing in S- or \umode, respectively.
This slows kernel boot-up by 0.44\%.

\paragraph{Enclave Lifecycle Costs.} 
Before performing larger-scale benchmarks, we assessed, what overheads can be expected under normal enclave execution.
Tab.~\ref{tab:overhead_minibench}~left, rows\,8-11 summarize measured overheads for executing helloworld enclave from Keystone.
Considering the overhead introduced for handling the \mmode timers, we observe an overhead for executing the enclave with 0.12\%, as expected.
Context switches into an enclave cause an overhead of 18\%, which is due to our saves and restore routine and the checking of the ecall. 
The overhead for enclave creation and deletion varies depending on the size of the loaded runtime binary.

\begin{table}[]
  \centering
  \caption{Summary of (left) Lifecycle Costs for Baseline vs. \codename and (right) Impact of \codename on RV8 when executing in U-mode and enclave.}
  \label{tab:overhead_minibench}
  \resizebox{0.37\columnwidth}{!}{%
  \begin{tabular}{@{}clr@{}}
  \toprule
  Stage                    & Component   & \%     \\ \midrule
  \multirow{3}{*}{\rotatebox[origin=c]{90}{Init}}    & OpenSBI     & 5.29   \\
                           & Monitor     & 91.28  \\
                           & M-Mode      & 20.81  \\ \midrule
  \multirow{3}{*}{\rotatebox[origin=c]{90}{Kernel}}  & Boot        & 0.44   \\
                           & Ecall       & 290.00 \\
                           & Timer       & 142.00 \\ \midrule
  \multirow{4}{*}{\rotatebox[origin=c]{90}{Enclave}} & Creation    & 15.58  \\
                           & Execution   & 0.12   \\
                           & Deletion    & 17.17  \\
                           & Ctx. Switch & 18.30  \\ \bottomrule
  \end{tabular}
  }
  \hspace{25pt}
  \resizebox{0.45\columnwidth}{!}{%
  \begin{tabular}{@{}lrr@{}}
  
  \toprule
  Target & U-mode & Enclave \\ 
  RV8 & \% & \% \\ \midrule
  aes    & 0.35    & 0.28    \\
  dst.   & 5.58    & 0.29    \\
  miniz  & 0.14    & 0.63    \\
  norx   & 0.12    & 0.29    \\
  prime  & 0.11    & 0.89    \\
  qsort  & 0.20    & 0.28    \\
  sha    & 0.17    & 0.29    \\ \bottomrule
  \end{tabular}
  }
  \end{table}

\paragraph{Benchmark 1: CPU (RV8).} 
We execute the RV8~\cite{rv8-bench} benchmarks natively on the host OS and in an enclave to compare the overheads for executing CPU-intensive applications. 
Tab.~\ref{tab:overhead_minibench}~(right) summarizes the results.
For execution inside an enclave, we separately analyze the overheads for enclave creation, runtime, and deletion. 
Enclave creation adds overheads of $\approx$12.5\% for all benchmarks, deletion of $\approx$10.7\%.
These one-time overheads are caused by our \codename implementation in the \primary since it handles these operations directly.
This overhead varies and is more significant for smaller binaries.
The reason is that creation and deletion have a constant part (creating/deleting necessary configuration structures) and a variable part (checking transferred page tables before starting and deleting reserved memory before yielding).
Since the RV8 binaries have similar sizes, we did not observe a large variation in the overheads.
For the host-OS, \codename adds an overhead on average of 0.2\%, excluding Dhrystone, which overheads with 5.6\%. We observe in the measurements that the execution for the baseline test for Dhrystone is discretely either $\approx$1.4B or $\approx$1.7B cycles.
For other applications, we observe an execution overhead <1\% for host-OS and in-enclave execution.

\begin{figure}[]
\centering
\begin{minipage}[c]{0.35\textwidth}
\includegraphics[width=1\linewidth]{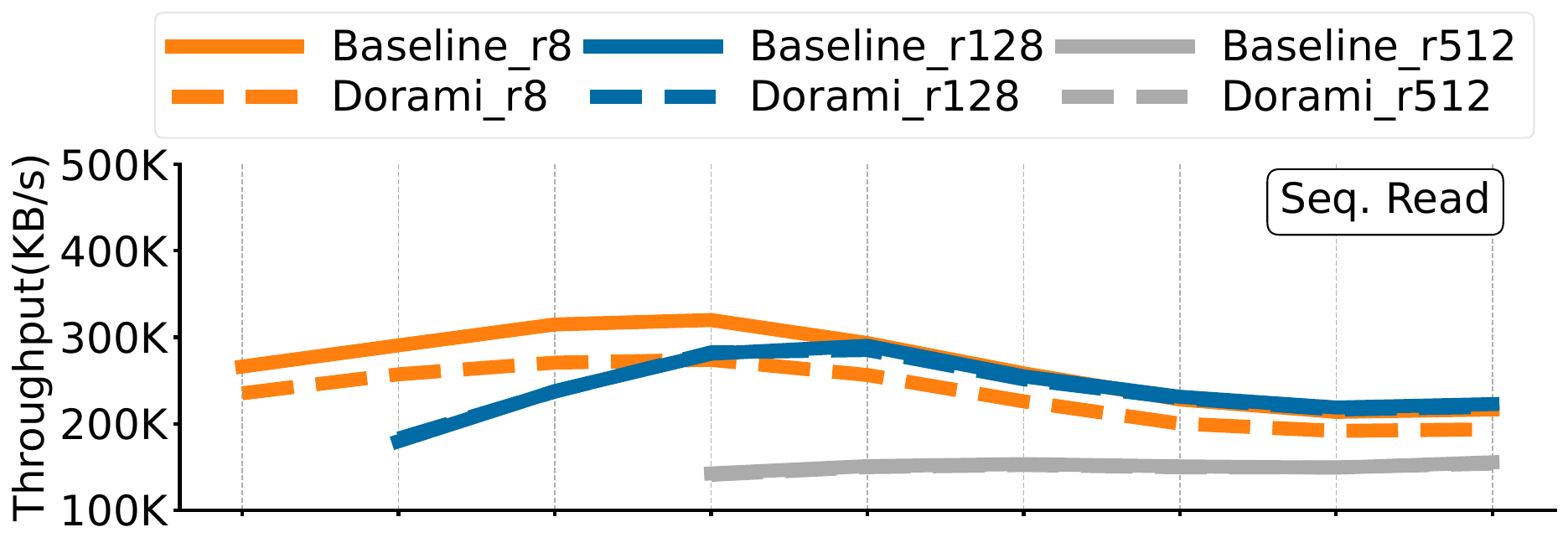}
\centering
\includegraphics[width=1\linewidth]{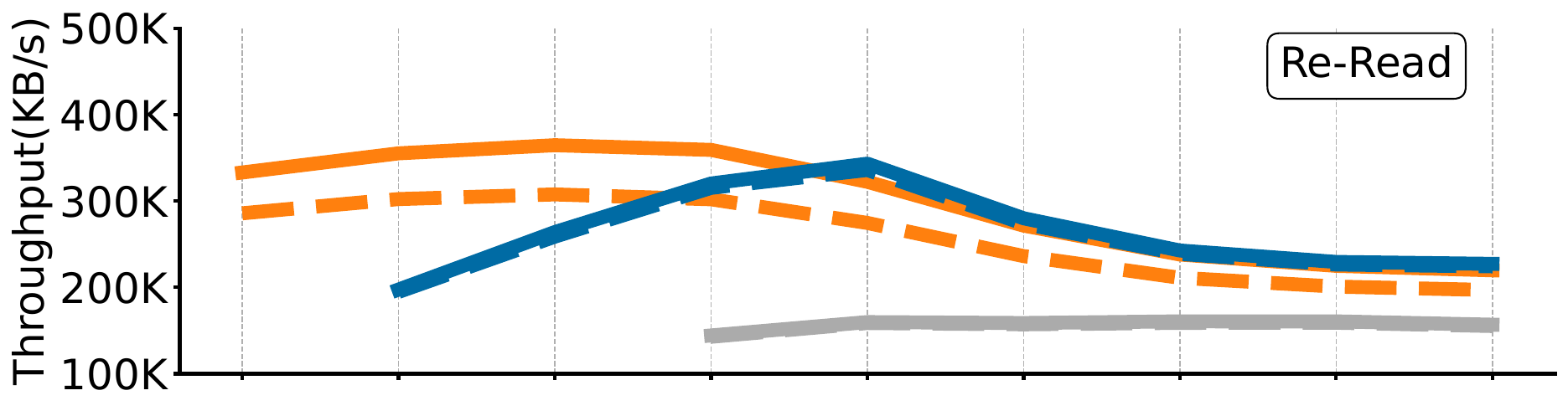}
\centering
\includegraphics[width=1\linewidth]{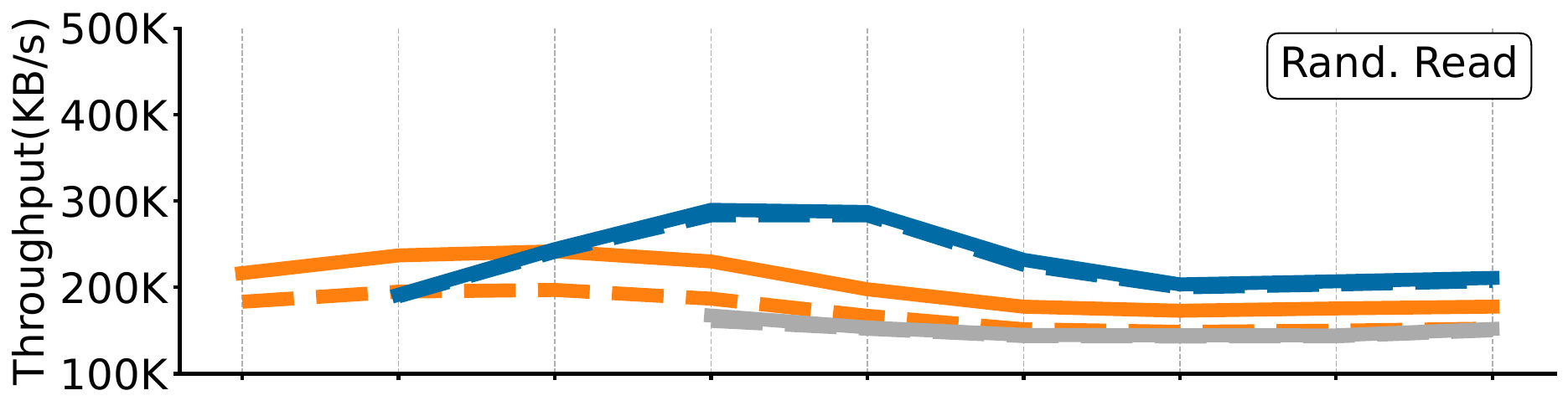}
\centering
\includegraphics[width=1\linewidth]{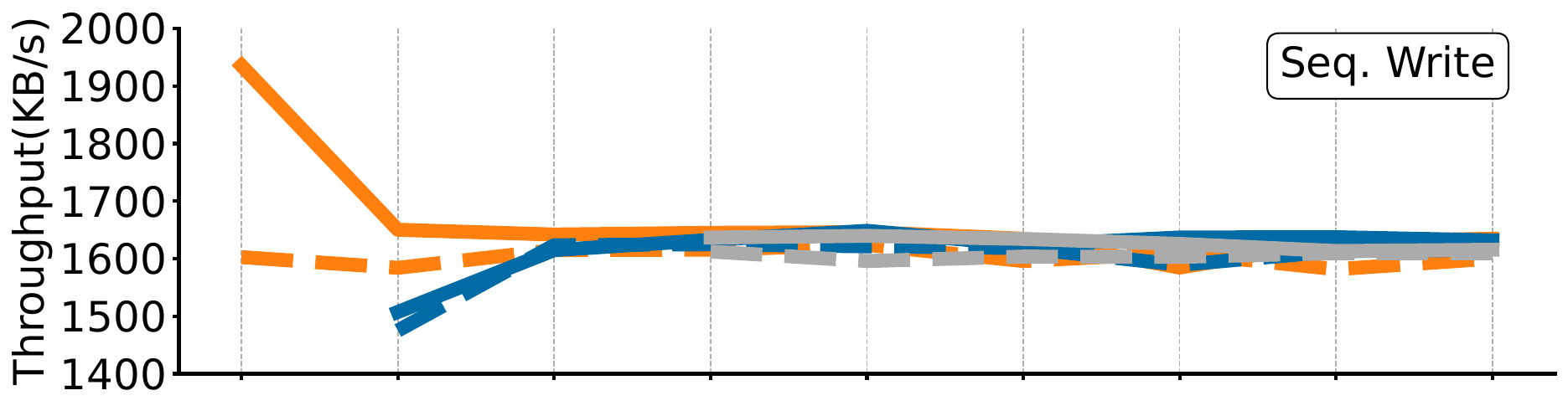}
\centering
\includegraphics[width=1\linewidth]{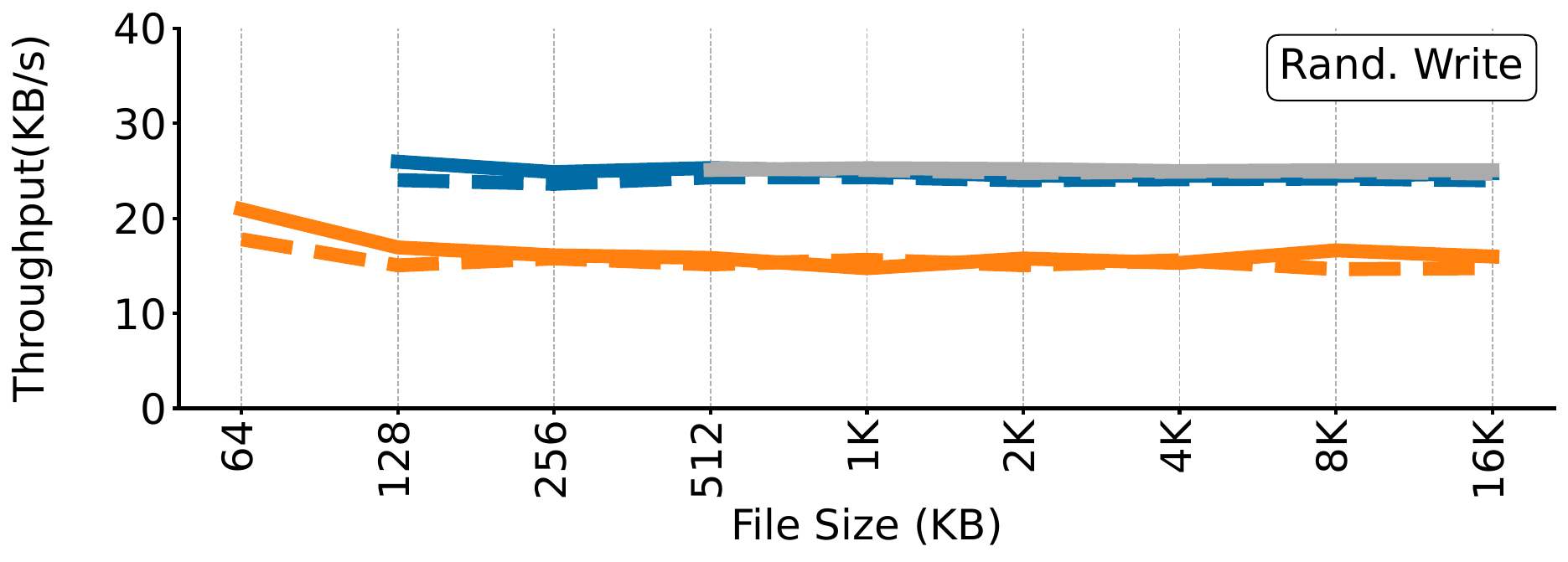}
\end{minipage}
\caption{Throughput report for FSCQ-largefiles.}
\label{fig:throughputgraphs}
\end{figure}

\paragraph{Benchmark 2: IO (FSCQ-Largefiles).} %
The RV8 benchmarks show enclave overheads induced by asynchronous context switches from scheduling.
However, I/O operations require synchronous exits (ocalls) to the host OS.
To measure \codename's impact on ocalls we use the FSCQ-largefile benchmark. 
It performs file I/O operations with file sizes from 8KB to 16MB and batch sizes of 8KB, 256KB, and 512KB.
It reads and writes to the files both sequentially and randomly.
Fig.~\ref{fig:throughputgraphs} shows the throughput in all experiments.
As expected, the throughput decreases across all experiments in the \codename setup.
Operations with smaller batch sizes show larger performance gaps due to more frequent context switches.
Random writes with smaller batches perform significantly worse, compared to larger batch sizes.
However for sequential writes, smaller batches initially perform better but eventually stabilize at around 1650 KB/s across all file sizes.
This is due to the 512KB block size of the file system, which requires the kernel to read before writing data in smaller batches.
This issue is particularly pronounced for random writes, where smaller batches are more likely to access uncached blocks.

\begin{figure}[]
\centering
\begin{subfigure}{0.49\linewidth}
		\includegraphics[width=\linewidth]{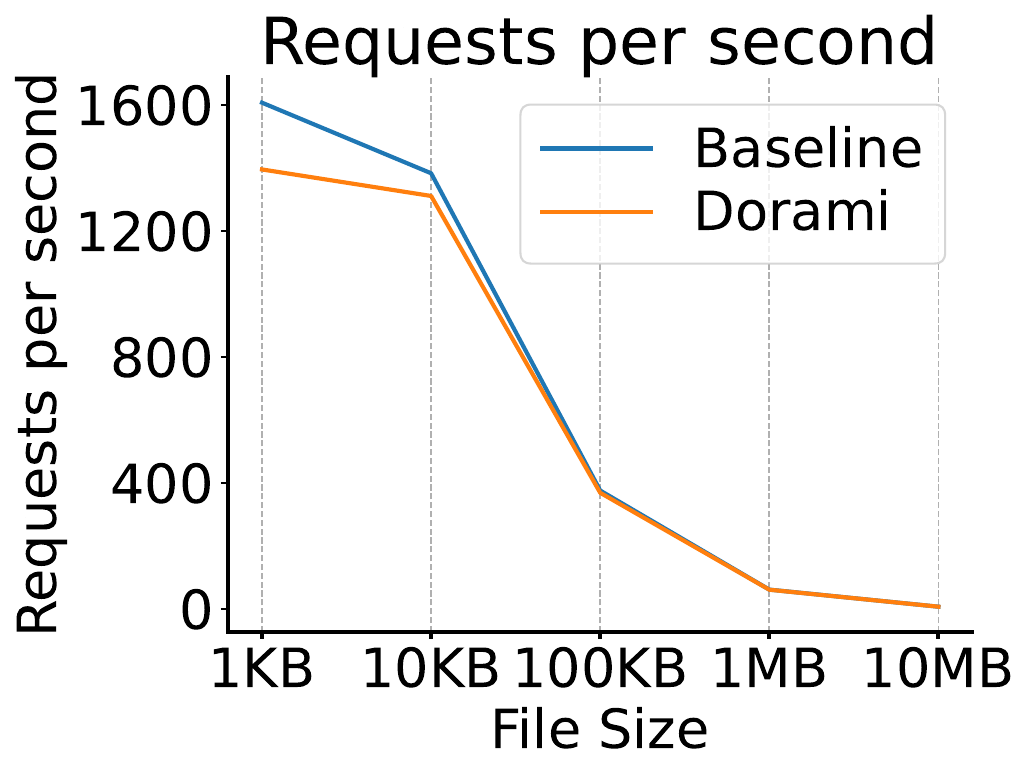}
\end{subfigure}
\centering
\begin{subfigure}{0.49\linewidth}
		\includegraphics[width=\linewidth]{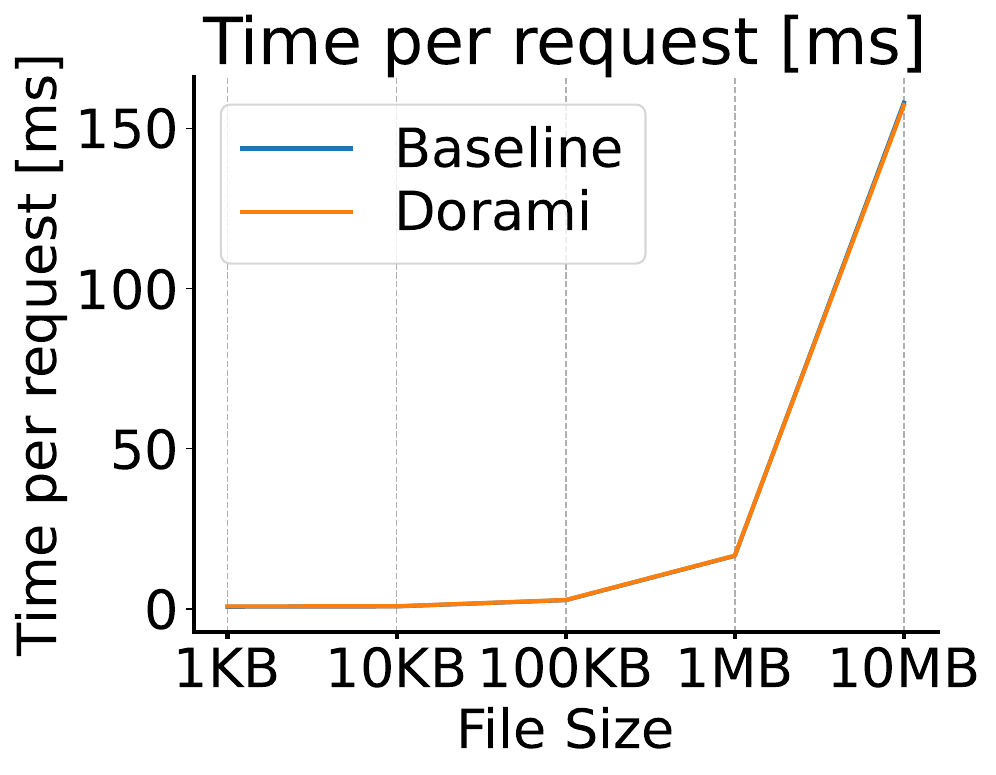}
\end{subfigure}
\caption{Darkhttpd: Requests per second and Time per request (latency) for Baseline and \codename, sizes 1\,KB-10\,MB.}
\label{fig:darkhttpd}
\end{figure}

\paragraph{Case-study 1: Webserver (darkhttpd).}
We demonstrate \codename's compatibility with real-world applications.
Our first case-study is Darkhttpd, a webserver to serve html-based webpages~\cite{darkhttpd}.
We take darkhttpd version from Cerberus which is compatible with Keystone~\cite{cerberus}.
It executes the webserver in an enclave and serves webpages to the network using ocalls.
We use apachebench to test webpages varying from sizes of 1KB to 10MB~\cite{apachebench}.
Fig.~\ref{fig:darkhttpd} shows the number of requests per seconds (RpS) and the time per request (TpR) for baseline (Keystone) and \codename for each file size. 
Only smaller file sizes show noticeable overheads, where RpS decreases by 13\% and TpR by 15\% for \codename.
For a larger sizes (e.g., 1\,MB), \codename overheads drop to 0.45\%. It is almost 0 for 10\,MB.
Since RpS and TpR differences are only significant for smaller files, we conclude that \codename would not significantly impact SLAs.
\codename does not impact ocalls (Row 3, Tab.~\ref{tab:darkhttpd}), the setup incurs 11 ocalls with 9 more for large file size to load more blocks.
\codename incurs a fixed overhead for \mmode enters and timer interrupts.
Tab.~\ref{tab:darkhttpd} shows that the total events increase with file size as expected---the larger the file, longer it takes to serve it, resulting in more timer interrupts. 
When we normalize the number of enters and timers per unit of file size (1KB), we see fewer events for larger files, which is expected---larger files spend more time on I/O while either yielding CPU time slices or blocking the CPU execution for IO. Importantly, as the normalized events decrease, \codename overhead reduces.

\begin{table}[]
\centering
\caption{darkhttpd event counts for baseline (B) and \codename (D) execution. ocalls (row 3), total and normalized M-mode enters (row 4-5) , timer interrupts (row 6-7). }
\label{tab:darkhttpd}
\resizebox{\columnwidth}{!}{%
\begin{tabular}{@{}llS[table-format=2.3]S[table-format=2.3]S[table-format=2.3]S[table-format=2.3]S[table-format=2.3]S[table-format=2.3]S[table-format=2.3]S[table-format=2.3]S[table-format=2.3]S[table-format=2.3]@{}}
\toprule
\multicolumn{2}{l}{\multirow{2}{*}{\textbf{Event}}} &
  \multicolumn{2}{c}{\textbf{1KB}} &
  \multicolumn{2}{c}{\textbf{10KB}} &
  \multicolumn{2}{c}{\textbf{100KB}} &
  \multicolumn{2}{c}{\textbf{1MB}} &
  \multicolumn{2}{c}{\textbf{10MB}} \\ \cmidrule(l){3-4} \cmidrule(l){5-6} \cmidrule(l){7-8} \cmidrule(l){9-10} \cmidrule(l){11-12}
\multicolumn{2}{l}{} &
  \multicolumn{1}{c}{\textbf{B}} &
  \multicolumn{1}{c}{\textbf{D}} &
  \multicolumn{1}{c}{\textbf{B}} &
  \multicolumn{1}{c}{\textbf{D}} &
  \multicolumn{1}{c}{\textbf{B}} &
  \multicolumn{1}{c}{\textbf{D}} &
  \multicolumn{1}{c}{\textbf{B}} &
  \multicolumn{1}{c}{\textbf{D}} &
  \multicolumn{1}{c}{\textbf{B}} &
  \multicolumn{1}{c}{\textbf{D}} \\ \midrule
Ocalls                    & Total & 11    & 11    & 11    & 11    & 11    & 11    & 11    & 11     & 20      & 20      \\
\multirow{2}{*}{M enters} & Total & 37.74 & 38.65 & 43.83 & 45.58 & 59.06 & 65.57 & 413   & 438.04 & 2686.83 & 3212.69 \\
                          & Norm  & 37.74 & 38.73 & 4.28  & 4.45  & 0.57  & 0.64  & 0.40  & 0.43   & 0.26    & 0.31    \\
\multirow{2}{*}{Timers}   & Total & 1.38  & 1.39  & 1.39  & 1.39  & 2.00  & 2.64  & 6.11  & 6.10   & 10.56   & 24.10    \\
                          & Norm  & 1.38  & 1.39  & 0.13  & 0.13  & 0.02  & 0.026 & 0.006 & 0.006  & 0.001   & 0.002    \\ \bottomrule
\end{tabular}%
}
\end{table}

\paragraph{Case-study 2: In-memory Database (SQLite).}
We measure a simple database server based on SQLite from Cerberus~\cite{cerberus}.
SQLite executes in an enclave and loads a database from the host using ocalls.
As in Cerberus, we measure 1,000 \verb|SELECT| queries and observe 5\,\% overhead compared to the baseline. 

\section{Discussion}
\label{sec:discussion}

We first discuss an approach for extending \codename from one to multiple F compartments.
Then we outline security considerations for implementing ePMP in hardware.

\subsection{Supporting Multiple F Compartments}\label{disc:ncompartments}
\begin{figure}
    \centering
    \includegraphics[width=1\columnwidth]{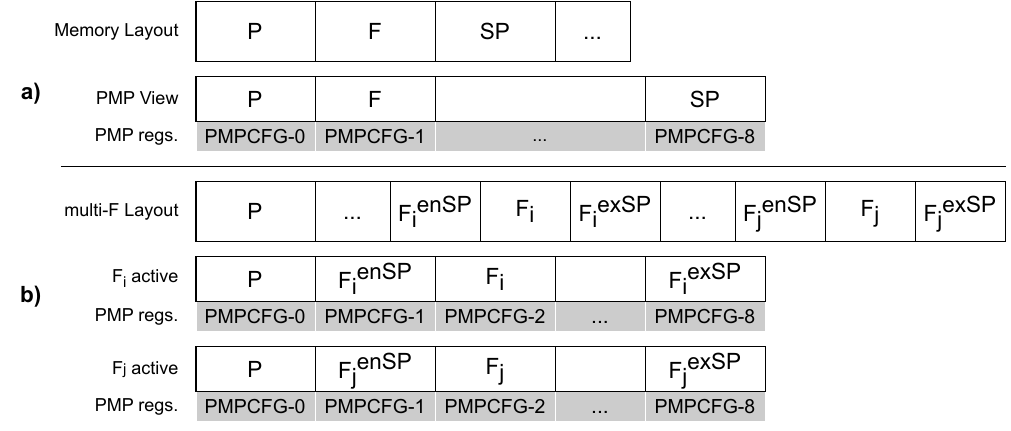}
    \caption{\textbf{multi-F: Memory layout and PMP configs}.\\ 
    \textbf{P}=\compi, \textbf{F}=\compii, \textbf{SP} =SallyPort, \textbf{enSP}=entry-SP; \textbf{exSP}=exit-SP.
    a) shows extract of \codename's memory layout and PMP config with one F-compartment. b) shows the same for a mullti-F implementation with 2 F compartments.}
    \label{fig:multi-F}
\end{figure}

\codename's design can be extended to create multiple F compartments, for example to isolate different firmware modules. 
Fig.~\ref{fig:multi-F}(a) shows the current memory layout of \codename for supporting one P and one F compartment with the corresponding PMP configurations. 
In original \codename design we placed the code section of \compii directly after the code section of \compi for compartment transitioning~(\ref{sec:p2s}).
However this approach can not apply as-is to $n$ compartments.
We can solve this issue by placing an SallyPort Entry (enSP) and an SallyPort Exit (exSP) before and after each of the n Firmware compartments. 
Fig.~\ref{fig:multi-F}(b) shows the memory layout for one P and multiple Fs.
In fact, such a solution would not even need more PMP entries per compartment i.e., the number of needed entries does not need to increase with the number of compartments.
Before transitioning into a particular F compartment (e.g., $F_j$), P can re-use existing entry for another F (e.g., $F_i$) and configure them with the $F_j$'s configuration.
This way, all other compartments automatically become inaccessible since they are not covered by any ePMP configuration.
In total, compared to one F design, such a solution would only require 1 additional PMP entry to be permanently reserved for all the F compartments.

\subsection{Microarchitectural Considerations}
\label{disc:microarch}
During our evaluation, we noticed several behavioral differences between an emulated, a synthesized, and a hardware implementation of RISC-V CPUs. 
The most critical observations we had regarding security are TOCTOU issues in the update routine of the PMP configurations. 
From a technical standpoint, it should be a legitimate operation that an instruction placed at the end of the PMP region locks the current region, unlocks the next that covers a memory region besides itself, and continues the execution from there (as explained in Section~\ref{ssec:transitions}). 
The enforcement of this rule change should be immediate.
However we observe that this needs to be carefully implemented into the CPU pipeline before the execution stage.
Let us take a look at the PMP implementation of the Rocketchip CPU. 
This CPU features a pipeline of 4-5 stages: instruction fetch+decode, execute, memory access, and writeback. 
When the CPU fetches and decodes an instruction regarding an update of the PMP configuration, the update in the register file only happens in the fourth cycle after the fetch in the writeback stage. 
However, the actual check against current PMP configurations happens directly at the instruction fetch phase. 
This means that there is a gap of 3 instructions that will be loaded into the pipeline before the fourth instruction fetch will actually consider updated PMP registers. 
Modern CPU designs such as Rocketchip and NOEL-V protect against this by marking such control-registers as hazardous.
This means that the pipeline is flushed and the instructions are re-fetched after the PMP instruction fully completes.
NOEL-V additionally needs to ensure that PMP-related hazards are synchronized on both pipelines when the Dual-Issue variant is in use.

Emulators such as Qemu check against PMP reconfigurations instantaneously while emulating the specific instruction.
Since Qemu does not implement a pipeline, it does not need to concern about hazards.

\section{Related Work}
\label{sec:related}

We discuss works beyond the ones covered in 
Section~\ref{sec:existing-approaches}.

\paragraph{Use of PMP in TEEs.}
Keystone~\cite{keystone} is one of the first TEEs on RISC-V that solely leverages PMP features without any hardware modifications. 
SPEAR-V, Elasticlave, and Cerberus also leverage PMP features to establish secure enclave systems~\cite{spearv, elasticlave, cerberus}.
Timber-V, Penglai, SERVAS, Cure, and Sanctum introduce hardware modifications to achieve memory isolation through alternative means, such as bus-level or CPU-component-based isolation~\cite{timberv, penglai, servas, cure, sanctum}.
While these TEE implementations assume the entirety of firmware to be trusted and error-free, \codename serves as a complementary addition rather than a competitive alternative. 

\paragraph{Confidential VMs.}
Modern TEEs offer a VM abstraction.
In addition, they 
introduce a new privilege level that executes trusted software beneath the VM. 
For example, Arm CCA executes a security monitor in EL3 as part of the trusted firmware, but also introduces a Realm Management Monitor (RMM) that executes in the EL2 of realm world to isolate different realm VMs~\cite{arm-da}.
Similarly, Intel TDX has a TD module that executes in a new privilege level called SEAM Root Mode~\cite{tdx}.
AMD SEV-SNP has Secure VM Service Module (SVSM) which optionally executes in VM Permission Level VMPL0~\cite{amd-sev}.
All of them use horizontal privilege separation. But the VM has to trust the guest kernel~\cite{veil-asplos2024}.

\paragraph{Extension of \codename to Different Architectures.}
Commercial products like Intel SGX and TDX, or AMD SEV-SNP, offer confidential computing capabilities through enclaves and VMs~\cite{costan2016intel, tdx, amd-sev}.
These platforms, feature fundamentally distinct memory isolation designs compared to RISC-V, implement proprietary measures for memory isolation, with closed-source firmware components. 
Similarly, Arm TrustZone~\cite{TZOS} and CCA~\cite{arm-da} introduce secure services and VM-based computing, respectively. 
TrustZone employs an Address Space Controller (ASC)~\cite{tzasc} for memory isolation, while CCA introduces Physical Address Space (PAS) regions. 
Arm encounters challenges in privilege separation of 
 code execution within the trusted firmware in EL3, equivalent to \mmode on RISC-V, which is 310 KLoC~\cite{cca-tf-a}.
Future works can investigate if \codename approach can be applied to CCA.

\paragraph{RISC-V Hypervisor Extensions.}
SMMTT~\cite{smmtt} isolates Physical Address Spaces for S-Mode software using memory tagging. 
SMMTT is a flexible alternative to Arm CCA~\cite{cove}, which also requires a universally trusted firmware in M-mode for configuration, similar to PMP. 
\codename complements  SMMTT, \primary can configure both SMMTT and ePMP, with remaining firmware in a separate compartment.

\section{Conclusion}
\codename is the first system that isolates the security monitor and the firmware on RISC-V. It uses existing standard ISA feature, ePMP, to achieve this goal with minimal overheads. 
\codename is compatible with current RISC-V firmware \& extensions and achieves reduction in the TCB. This can pave path for future works to formally verify the security monitor without reasoning about the rest of the firmware that executes alongside in the highest execution privileged layer on RISC-V.

\section*{Acknowledgments}
We thank the Usenix Security 2024 reviewers for 
pointing us to NOEL-V and for their constructive feedback that significantly improved the paper. 
Thanks to Laurent Wirz for analyzing OpenSBI to identify the partition boundaries, Mélisande Zonta-Roudes for feedback on the early version of the paper, and Supraja Sridhara for fruitful discussions on RISC-V TEEs.
This work was supported by the Swiss Joint Research Center financed by Microsoft Research. 

\balance
\bibliographystyle{plain}
\bibliography{references}

\end{document}